\documentclass[proof]{pasj00}
\RelativeFigurePath{./FIGS}
\color{black}

\newcommand{\fluxunit}{erg~cm$^{-2}$~s$^{-1}$}
\newcommand{\sbunit}{erg~cm$^{-2}$~s$^{-1}$~sr$^{-1}$}

\begin{document}
\Received{$\langle$reception date$\rangle$}
\Accepted{$\langle$acception date$\rangle$}
\Published{$\langle$publication date$\rangle$}
\SetRunningHead{A. Hoshino et al.}{X-ray Temperature and Mass Measurements to the Virial Radius of A1413 with Suzaku}
\title{X-ray Temperature and Mass Measurements to the Virial Radius of Abell 1413 with Suzaku
}
\author{
 A. Hoshino,\altaffilmark{1}
 J. P. Henry,\altaffilmark{2}
 K. Sato,\altaffilmark{3}
 H. Akamatsu,\altaffilmark{1}
 W. Yokota,\altaffilmark{1}
 S. Sasaki,\altaffilmark{1}
 Y. Ishisaki,\altaffilmark{1}
 T. Ohashi,\altaffilmark{1}\\
 M. Bautz,\altaffilmark{4}
 Y. Fukazawa,\altaffilmark{5}
 N. Kawano,\altaffilmark{5}
 A. Furuzawa,\altaffilmark{6}
 K. Hayashida,\altaffilmark{7}
 T. Tawa,\altaffilmark{7}
 J. Hughes,\altaffilmark{8}\\
 M. Kokubun,\altaffilmark{9}
and T. Tamura\altaffilmark{9} 
}
\altaffiltext{1}{
Department of Physics, Tokyo Metropolitan University,\\
 1-1 Minami-Osawa, Hachioji, Tokyo 192-0397}
\email{h\_akio@phys.metro-u.ac.jp}
\altaffiltext{2}{
Institute for Astronomy, University of Hawaii,\\
 2680 Woodlawn Drive, Honolulu, HI 96822, USA}
\altaffiltext{3}{
Graduate School of Natural Science and Technology,\\
 Kanazawa University, Kakuma, Kanazawa, Ishikawa 920-1192}
\altaffiltext{4}{
Kavli Institute for Astrophysics and Space Research, Massachusetts Institute of Technology,\\
77 Massachusetts Avenue, Cambridge, MA 02139, USA}
\altaffiltext{5}{
Department of Physical Science, Hiroshima University,\\
 1-3-1 Kagamiyama, Higashi-Hiroshima, Hiroshima 739-8526}
\altaffiltext{6}{
EcoTopia Science Institute and Department of Astrophysics, Nagoya University,\\
Furo-cho, Chikusa-ku, Nagoya 464-8603}
\altaffiltext{7}{
Department of Earth and Space Science, Osaka University,\\
 Toyonaka, Osaka 560-0043}
\altaffiltext{8}{
Department of Physics and Astronomy, Rutgers University\\
 136 Frelinghuysen Road, Piscataway, NJ 08854-8019, USA}
\altaffiltext{9}{
Department of High Energy Astrophysics, Institute of Space and 
Astronautical Science,\\ Japan Aerospace Exploration Agency,\\
 3-1-1 Yoshinodai,Sagamihara, Kanagawa 229-8510}
\maketitle
\begin{abstract}
We present X-ray observations of the northern outskirts of the
relaxed galaxy cluster A1413 with Suzaku, whose XIS instrument has
the low intrinsic background needed to make measurements of these
low surface brightness regions. We excise 15 point sources
superimposed on the image above a flux of $1\times 10^{-14}$
\fluxunit\ (2--10~keV) using XMM-Newton and Suzaku images
of the cluster. We quantify all known systematic errors as part of our
analysis, and show our statistical errors encompasses them for the
most part.  Our results extend previous measurements with Chandra
and XMM-Newton, and show a significant temperature drop to about 3~keV
at the virial radius, $r_{200}$.  Our entropy profile in the outer region
($> 0.5\; r_{200}$) joins smoothly onto that of XMM-Newton,
and shows a flatter slope compared with simple models,
similar to a few other clusters observed at the virial radius.
The integrated mass of the cluster at the virial radius is approximately
$7.5\times10^{14}M_{\odot}$ and varies by about 30\% depending on
the particular method used to measure it.
\end{abstract}
\section{Introduction}
X-ray observations of intracluster medium (ICM) primarily give density
and temperature. Density may be deduced relatively straightforwardly
from cluster surface brightness because the ICM is optically thin and
the emission coefficient over most observed bandpasses is nearly
independent of temperature. There is good agreement on the ICM density
over the observed radial range among different observers. In contrast,
cluster temperatures have not been measured much beyond about half of
the virial radius and, until recently, the shape of the temperature
radial profile was a matter of heated debate even to that radius. Now
independent measurements using four different observatories are
consistent with a factor of $\sim2$ decline of the projected
temperature from the center to half the virial radius, at least for
relaxed clusters \citep{markevitch98, degrandi02, vikhlinin05,
  piffaretti05, pratt07}.
 
In addition to the fundamental density and temperature variables, it is
possible to derive additional thermodynamic variables from them, such
as pressure and entropy. These derived variables are very useful when
trying to understand the gravitational and non gravitational processes
that were operative during the formation and evolution of the
cluster. With the assumption of hydrostatic equilibrium, the cluster's
total mass can also be derived from the ICM temperature and the radial
derivatives of temperature and density. While this assumption is not
valid for many clusters, X-ray observations are one of the few that
can be used to measure such masses.

The cold dark matter (CDM) paradigm combined with numerical
simulations predict that the structure of clusters should exhibit
self-similar scaling. That is their properties should be the same when
scaled appropriately by redshift and the virial radius. This expected
behavior occurs because clusters form from scale-free density
perturbations and their evolution is mainly set by scale-free gravity,
and both of these result because cluster masses are mainly
CDM\@. Among the two most studied scaling relations are the radial
profiles of temperature and total mass. The mass profile in particular
is named the NFW profile after the authors of one of the original
papers on this subject \citep{navarro96}. Deviations from the expected
scaling in high-quality data indicate the importance of non
gravitational processes and/or unaccounted bias in the data.  The
temperature profile predicted by numerical simulations shows a
significant drop with radius to about one third of the peak value at
the virial radius (e.g.\
\cite{loken02,komatsu02,borgani04,roncarelli06}). Observationally, the
temperature profiles are the key factor in deriving the cluster mass
profile up to the virial radius. The precise mass profile will allow
us to judge the validity of the present CDM framework, and gives
assurance for our application of cluster properties to cosmological
studies.

For all of the above reasons it is important to extend X-ray cluster
temperature measurements beyond the current limit of $\sim0.5$ of the
virial radius, particularly for relaxed clusters. Suzaku observed
several such clusters in these regions, and some of the results have
been published including PKS0745$-$191 \citep{george08}, A1795
\citep{bautz09}, and A2204 \citep{reiprich09}. All these clusters show
a systematic trend of the temperature dropping to about one third of
the central value, broadly consistent with theoretical expectations.
However the statistical quality of the data for any individual cluster
is limited and we need to look at many others to discern general
behavior.  The main difficulty for the measurement of ICM properties
in the virial region is the low cluster surface brightness, which
means that in no energy range does the cluster emission exceed the
Galactic foreground plus cosmic X-ray background emission. Careful
study of systematic errors is therefore mandatory when trying to
assess the ICM properties around the virial radius.

We have made Suzaku observations of A1413, a moderately distant
cluster at redshift $z=0.1427$ \citep{boehringer00} whose size is well suited
to our field of view. Assuming a Hubble constant of 70
km~s$^{-1}$~Mpc$^{-1}$ or $h_{70}=1$ as well as cosmological parameters
of $\Omega_\mathrm{m0} = 0.28$ and $\Omega_{\Lambda0} = 0.72$,
we imply an angular diameter distance of
$519\; h_{70}^{-1}$ Mpc, a luminosity distance of $678\; h_{70}^{-1}$
Mpc and a scale of $151.2\; h_{70}^{-1}$~kpc per arcmin. 

Although we will measure this cluster's properties at the virial
radius, we must make rather coarse spatial bins to do so. Thus we can
not measure the actual virial radius with much precision. As a point
of reference we adopt an often used nominal value, $r_{200}$, which is
the radius within which the cluster average density is 200 times the
critical density needed to halt the expansion of the universe. For our
cosmology
\begin{eqnarray}
r_{200} = 2.59\; h^{-1}_{70} \sqrt{\langle kT\rangle /10 ~{\rm keV}}~{\rm Mpc},
\end{eqnarray}
in which $\langle kT\rangle$ is the cluster average temperature
\citep{2009ApJ...691.1307H}. An overdensity of 200 is contained within
the virialized region of a spherical collapse in an Einstein-de Sitter
universe at all red-shifts. Generalizing to a spherical collapse for
our adopted cosmology at the red-shift of A1413, gives an overdensity
of 109 for the virialized region \citep{henry00}. However $r_{109}$ is
only 22\% larger than $r_{200}$.  So for comparison with previous work
we adopt the latter as the nominal virial radius. 

The average temperature of A1413 integrated over the radial range of
$70$ kpc to $r_{500}$ is $7.38 \pm 0.11$ keV
\citep{vikhlinin06}, where $r_{500}$ is defined analogously to
$r_{200}$, implying $r_{200} = 2.24\; h^{-1}_{70}$ Mpc or $14'.8$.
Previous observations indicate the cluster is relaxed and there are
high quality temperature and mass radial profiles available from both
XMM-Newton and Chandra \citep{pointecouteau05, vikhlinin06}.  There is
some disagreement about the mass profile of A1413 in these two works.
\citet{pointecouteau05} find $r_{500} = (1.13 \pm 0.03)\; h^{-1}_{70}$
Mpc and $M_{500} = (4.82 \pm 0.42) \times 10^{14}\; h^{-1}_{70}\; M_{\odot}$,
while \citet{vikhlinin06} find $(1.34 \pm 0.04)\; h^{-1}_{70}$ Mpc and
$(7.79 \pm 0.78) \times 10^{14}\; h^{-1}_{70}\; M_{\odot}$, respectively,
where $M_{500}$ is the mass within $r_{500}$.
Note that both observations measure the temperature out to
$r_{500}$ so the disagreement is not due to uncertainties in extrapolation.

Throughout this paper,
errors are at 90\% confidence for one interesting parameter
otherwise noted.

\section{Observations}

\subsection{Suzaku}

\begin{figure*}[htbp]
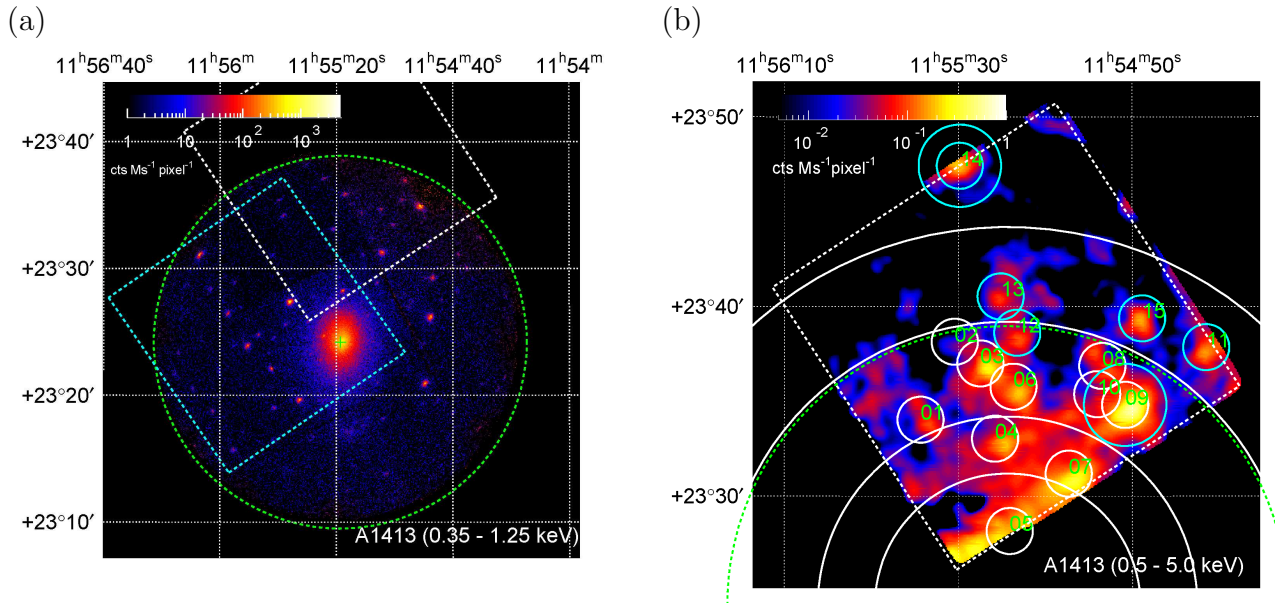

\begin{center}
\begin{minipage}{\textwidth}
\parbox{0.48\textwidth}{\ (a)}\hfill
\parbox{0.48\textwidth}{\ (b)}
\vspace*{-1.5ex}
\end{minipage}
\begin{minipage}{\textwidth}
\FigureFile(0.48\textwidth,\textwidth){figure1a.eps}\hfill
\FigureFile(0.48\textwidth,\textwidth){figure1b.eps}
\end{minipage}
\end{center}
\caption{ (a) XMM-Newton MOS1 + MOS2 image of A1413 in the
0.35--1.25~keV band.  The image is corrected for exposure, vignetting
and background. The white and blue boxes show the fields of view of
the Suzaku XIS\@ and Chandra ACIS \citep{vikhlinin06}.  The green
circle shows $r_{200}$ of $14'.8$. Color scale unit is
cts~Ms$^{-1}$~pixel$^{-1}$ (1 pixel = $2''.49\times2''.49$). (b)
Background subtracted Suzaku FI+BI image of the outskirts of A1413 in
the 0.5--5 keV band smoothed by a 2-dimensional gaussian with $\sigma
= 16''$. The image is corrected for exposure time but not for
vignetting. Color scale unit is cts~Ms$^{-1}$~pixel$^{-2}$ (1 pixel =
$1''.04\times 1''.04$). COR2 $>8$~GV and $100 <$ PINUD $<300$
cts~s$^{-1}$ screening was applied. The $^{55}$Fe calibration source
regions are also included in the figure, because they have negligible
counts in this energy band.  Large white circles denote $7'$, $10'$,
$15'$, and $20'$ from the surface brightness peak of the XMM-Newton
image. Small white and blue circles show the excluded point sources.  }
\label{fig:image}
\end{figure*}

We observed the northern region of A1413 
with the Suzaku XIS detectors. In table~\ref{tab:obslog}, we give the
details of our observation, and in figure~\ref{fig:image}(a), we show the
XIS field of view (FOV) superimposed on the XMM-Newton image of A1413. The
XIS instrument consists of 4 CCD chips; one back-illuminated (BI: XIS1)
and three front-illuminated (FI: XIS0, XIS2, XIS3), with each is combined
with an X-ray telescope (XRT). The IR/UV blocking
filters had accumulated a significant contamination by the time of the
observation since its launch (July 2005);
we include its effects on the effective area in our analysis.
The XIS was operated with normal clocking mode, in
$5\times 5$ or $3\times 3$ editing modes.
The spaced-row charge injection (SCI) was not applied,
and all the four CCDs were working at the time of the observation.

We show the FI+BI image in the 0.5--5~keV energy band in
figure~\ref{fig:image}(b). The non X-ray background (NXB), cosmic X-ray
background (CXB), and the Galactic background components (GAL) are
subtracted as described below, and the result smoothed by
a 2-dimensional gaussian with $\sigma=16''$ are shown.
The image is corrected for
exposure time variations, but not for vignetting. Screening
requirements are COR2 $>8$ GV and $100 < {\rm PINUD} < 300$ cts~s$^{-1}$,
where COR2 is the cut-off-rigidity calculated with the most recent
geomagnetic coordinates and PINUD is the count rate from the upper
level discriminatory of the Hard X-ray Detector (HXD) PIN silicon diode
detectors (see \cite{tawa08}). The circles with $70''$ and $125''$
radii enclose excluded point sources. The small white circles indicate
point sources detected in the XMM-Newton data. Blue circles show
sources selected by eye in the Suzaku image. 

We used HEAsoft ver 6.4.1 and CALDB 2008-06-21 for all the Suzaku
analysis presented here.  We extracted pulse-height spectra in five
annular regions from the XIS event files. The inner and outer radii of
the regions were $2'.7-7'$, $7'-10'$, $10'-15'$, $15'-20'$, and $20'-26'$,
respectively, measured from the
XMM-Newton surface brightness peak of A1413 at (R.A., Dec.) =
(\timeform{11h55m18.7s}, \timeform{23\deg01'48''}) in J2000.
We analyzed the spectra in the 0.5--10~keV range for the FI detectors
and 0.4--10~keV for the BI detector.
In the the $2'.7-7'$ annulus,
we ignored the energy band 5--7~keV for the FI detectors
when we analyzed the spectra, because those data were affected by
Mn-K$_\alpha$ (5.9~keV) X-rays from the $^{55}$Fe calibration source.
In other annuli, positions of the calibration sources themselves
were masked out using the {\it calmask} calibration database (CALDB) file.

\begin{table*}[bt]
\caption{
Log of Suzaku observations of Abell 1413
}\label{tab:obslog}
\begin{center}
\begin{tabular}{lll}
\hline\hline
\makebox[0.46\textwidth][l]{Observation ID $\dotfill$} & 800001010  \\
Date of observation $\dotfill$ & 2005-Nov-15 19:54:46 -- 2005-Nov-18 14:14:45  \\
Exposure time (ks) \\
~~(COR2 $>$ 0 GV) $\dotfill$ &
  XIS0: \makebox[0in][r]{1}07.4, XIS1: \makebox[0in][r]{1}08.0, XIS2: \makebox[0in][r]{1}07.5, XIS3: \makebox[0in][r]{1}07.6\\
~~(COR2 $>$ 8 GV) $\dotfill$ &
  XIS0: 76.1, XIS1: 76.4, XIS2: 76.2, XIS3: 76.2\\
~~(COR2 $>$ 8 GV and 100 $<$ PINUD $<300$ cts~s$^{-1}$) $\dotfill$ &
  XIS0: 71.9, XIS1: 72.0, XIS2: 72.0, XIS3: 72.0\\
(R.A., Dec.) in J2000 $^{\ast}$ $\dotfill$ & (\timeform{11h55m19.0s}, \timeform{23D24'30''})\\
XIS mode $\dotfill$ & $5\times5\;/\;3\times3$, normal clocking, window off, SCI off \\
$N_{\rm H}$ $\dotfill$ &  $2.19\times10^{20}$ cm$^{-2}$ \citep{dickey90}\\
\hline\\[-1ex]
\multicolumn{3}{l}{\parbox{0.97\textwidth}{\footnotesize\footnotemark[$\ast$]
Average pointing direction of the XIS, shown by the RA\_NOM  
and DEC\_NOM keywords of the FITS event files.
   }}\\
\end{tabular}
\end{center}
\end{table*}

\subsection{XMM-Newton}
We analyzed an image in the energy band 0.35--1.25~keV taken with
XMM-Newton \citep{pratt02}. This observation was carried out in June
2000 (OBSID: 0112230501). The exposure time was 25.7 ks
(MOS1,MOS2). SAS ver 6.0 and HEAsoft ver 6.4.1 were used for the
analysis. XMM-Newton has much higher spatial resolution compared to
Suzaku. We used this image as input for the response simulators and to
find point sources.  \citet{pratt02} derived a ratio of minor to major
axis to be 0.71 and a position angle $2^\circ 26'$ based on the XMM
data. Since Suzaku coverage is limited in the north section of the
cluster as shown in figure \ref{fig:image}, we did not include the
cluster ellipticity in our analysis.


\section{Background Analysis}
Accurate estimation of the background is particularly important when
constraining the ICM surface brightness and temperature in the outer
region of clusters. We assumed that the background is comprised of
three components: non-X-ray background (NXB), cosmic X-ray background
(CXB) and Galactic emission (GAL), which itself is comprised of two
components. In this section we describe how we estimate all these
background components.

\subsection{Point Source Analysis}

We want to excise point sources because we are only interested in this
paper in the ICM. However, since the CXB is comprised of faint point sources,
we then need to correct the background level for the resolved sources. This
and the next section describe the procedure we used for these tasks.

\begin{figure*}[htbp]
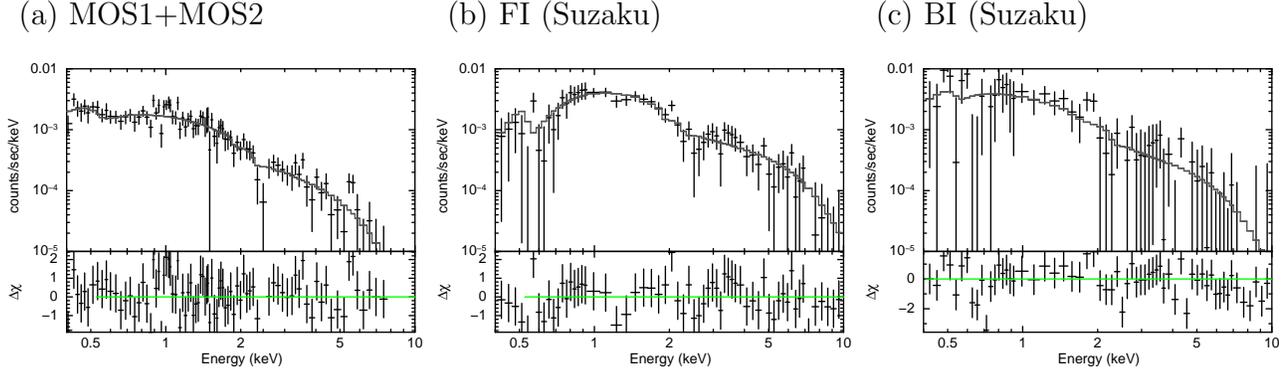

\begin{minipage}{\textwidth}
\parbox{0.33\textwidth}{\ (a) MOS1+MOS2}\hfill
\parbox{0.33\textwidth}{\ (b) FI (Suzaku)}\hfill
\parbox{0.33\textwidth}{\ (c) BI (Suzaku)}
\end{minipage}
\begin{minipage}{\textwidth}
\FigureFile(0.33\textwidth,\textwidth){figure2a.eps}\hfill
\FigureFile(0.33\textwidth,\textwidth){figure2b.eps}\hfill
\FigureFile(0.33\textwidth,\textwidth){figure2c.eps}
\end{minipage}
\vspace*{-2ex}
\caption{
Power-law model fit to the sum of all point source spectra.
(a) MOS1+MOS2, (b) FI, and (c) BI
(black: source spectra, grey: best-fit model).
}\label{fig:ps-spec}
\end{figure*}

We used the XMM-Newton image to detect point sources in the XIS FOV
because its spatial resolution ($14''$ half power diameter; HPD)
is better than Suzaku's ($2'$ HPD)\@.
We detected 10 point sources using {\it wavdetect} of CIAO,
and extracted source and background spectra by setting
the extraction radius of $33''$ and $33''-66''$, respectively.
First, we checked that the MOS1 and MOS2 spectra of each source
were consistent. Then we summed the MOS1 and MOS2 spectra
to increase the statistics, and fitted the spectrum of
each source to evaluate individual spectral parameters.
Finally we added the spectra of all the point sources to
estimate how much of the CXB these sources resolve.
We fitted the spectra by ${\it wabs}\times {\it pegpwrlw}$.
The best-fit parameters for the individual point sources and
their sum are shown in table~\ref{tab:pointsource}.
We obtained $\chi^2$/dof = 87.2/77 for the
power-law fit to the combined spectrum (figure~\ref{fig:ps-spec}(a)),
indicating a reasonable spectral fit.
The photon index is $\Gamma=1.92\pm 0.09$ and
the flux is $3.23^{+0.48}_{-0.44}\times 10^{-13}$ \fluxunit.

We also searched for point sources located outside of the XMM-Newton
field with Suzaku, finding an additional 5 sources by eye. These
sources all show statistical significance higher than $3.9\sigma$
against the brightness fluctuation around individual sources.
We performed spectral fits to all the point sources with
Suzaku according to the following procedure.
The source photons came from a circle of $40''$ radius with
encircling annular background region of $40''-100''$ radii.
We selected the source regions so they did not overlap each other.
These source and background areas could be slightly different
among the detectors and sources due to filtering by the {\it calmask}
regions and the presence of hot pixels.
We added the FI spectra from XIS0, XIS2, and XIS3 detectors,
and summed the BACKSCAL keyword in the FITS header,
which correspond to the area of extraction region,
$A_{\rm sr}$ or $A_{\rm bg}$.
Then, we carried out spectral fits for the FI and BI spectra
simultaneously using the same spectral model as before,
first for the individual sources and then the sum of all the point sources.

We show the best-fit parameters for the individual point sources and
their sum in table~\ref{tab:pointsource} except for sources 04, 05,
and 08, because they were faint so that we could not estimate
their background reasonably.
Obtained fluxes of the sources are slightly affected by
leaked photons of the target to the surrounding background regions.
To correct for this effect, we calculated the ratio, $f_{\rm leak}$,
of the leaked photons
in each background region to the detected photons in the source region
using the ``{\it xissim}'' FTOOL \citep{ishisaki07}.
We corrected the original source flux by multiplying a factor
$1/(1-f_{\rm leak}\; A_{\rm sr} / A_{\rm bg})\simeq 1/(1-0.2\; f_{\rm leak})$
in the $F_{\rm X}$ columns of Suzaku in table~\ref{tab:pointsource}.
Figures~\ref{fig:ps-spec}(b) and (c) show the combined spectra of
all sources for FI and BI\@.
We obtained $\chi^2$/dof = 113.1/117 for
the power-law fit to the combined spectrum, indicating a reasonable
spectral fit, too. The photon index is $\Gamma=1.82\pm0.12$ and the flux is
$4.83^{+0.60}_{-0.56}\times 10^{-13}$ \fluxunit\ (2--10~keV)\@.

The number of sources we found and their total flux
are consistent with that expected from the $\log N$--$\log S$ relation
summarized in figure~20 of \citet{kushino02}.
The detected sources ranges from $\sim 10^{-14}$ to $\sim 10^{-13}$
\fluxunit.
We excised all the point sources detected in either the Suzaku or
XMM-Newton observations. Normally we excluded a region of $70''$
radius but used $125''$ radius for two sources (09 and 14 in
table~\ref{tab:pointsource}).

\begin{table*}[bt]
\caption{
Best-fit parameters of detected point sources.
}\label{tab:pointsource}
\begin{center}
\begin{tabular}{lccrccccrc}
\hline\hline \makebox[0in][l]{Source} & & \multicolumn{4}{c}{FI+BI (Suzaku)\makebox[0in][l]{\,$^\ast$}} && \multicolumn{3}{c}{MOS1+MOS2 (XMM-Newton)\makebox[0in][l]{\,$^{\dag}$}}\\
\cline{3-6} \cline{8-10}
\makebox[0in][l]{ID} & (R.A., Dec.) in J2000\hspace*{-0.5em} & $\Gamma $ &  \multicolumn{1}{c}{$F_{\rm x}$\makebox[0in][l]{\,$^\ddag$}} & $\chi^2$/dof & $f_{\rm leak}$ && $\Gamma$ &  \multicolumn{1}{c}{$F_{\rm x}$\makebox[0in][l]{\,$^\ddag$}} & $\chi^2$/dof \\
\hline
01 &(\timeform{11h55m38.7s}, \timeform{23D34'02''}) & $2.5_{-0.8}^{+1.3}$ & $<2.4$ & 28.0/18 & 1.53 &&		$1.9_{-0.4}^{+0.5}$ & $2.4_{-1.2}^{+1.8}$ & 16.6/17 \\ 
02 &(\timeform{11h55m30.8s}, \timeform{23D38'09''}) & \makebox[0in][c]{1.7 (fixed)} & $<{2.1}$ & 14.7/12 & 1.41 &&			$1.9_{-0.5}^{+0.6}$ & $2.3_{-1.4}^{+2.3}$ & 5.7/12 \\ 
03 &(\timeform{11h55m24.9s}, \timeform{23D37'00''}) & $2.2_{-0.4}^{+0.5}$ & $1.6_{-0.6}^{+0.8}$ & 37.4/30 & 1.34 &&	$1.7_{-0.5}^{+0.6}$ & $1.6_{-1.0}^{+1.6}$ & 11.2/9 \\   
04 &(\timeform{11h55m21.6s}, \timeform{23D33'02''}) & & & & && 								$1.8_{-0.3}^{+0.3}$ & $3.6_{-1.3}^{+1.6}$ & 11.1/24 \\   
05 &(\timeform{11h55m18.2s}, \timeform{23D28'10''}) & & & & && 								$1.1_{-0.4}^{+0.4}$ & $5.9_{-2.4}^{+2.9}$ & 63.6/58 \\    
06 &(\timeform{11h55m17.3s}, \timeform{23D35'47''}) & $1.4_{-0.4}^{+0.4}$ & $4.2_{-1.4}^{+1.6}$ & 28.1/27 & 1.39 &&	$2.0_{-0.2}^{+0.2}$ & $5.0_{-1.4}^{+1.7}$ & 28.2/28 \\ 
07 &(\timeform{11h55m04.6s}, \timeform{23D31'11''}) & $1.9_{-0.3}^{+0.3}$ & $4.7_{-1.9}^{+2.1}$ & 39.2/37 & 1.26 &&	$2.1_{-0.2}^{+0.2}$ & $4.1_{-1.0}^{+1.2}$ & 40.9/37 \\ 
08 &(\timeform{11h54m56.9s}, \timeform{23D36'52''}) & & & & &&								$1.3_{-0.5}^{+0.5}$ & $6.8_{-3.1}^{+4.4}$ & 5.4/5 \\ 
09 &(\timeform{11h54m51.7s}, \timeform{23D34'49''}) & $1.7_{-0.1}^{+0.2}$ & $\makebox[0in][r]{1}6.0_{-2.4}^{+2.6}$ & 79.1/58 & 1.43 &&	$2.2_{-0.2}^{+0.2}$ & $8.1_{-2.1}^{+2.4}$ & 42.5/41 \\ 
10 &(\timeform{11h54m58.1s}, \timeform{23D35'23''}) & \makebox[0in][c]{1.7 (fixed)} & $<1.8$ & 18.0/17 & 1.21 &&			$1.7_{-1.2}^{+1.3}$ & $1.1_{-1.1}^{+2.9}$ & 5.1/7 \\   
11 &(\timeform{11h54m33.0s}, \timeform{23D37'51''}) & $1.7_{-0.5}^{+0.6}$ & $3.1_{-1.9}^{+2.3}$ & 13.5/13 & 1.24 && & & \\   
12 &(\timeform{11h55m16.5s}, \timeform{23D38'37''}) & $2.0_{-0.5}^{+0.7}$ & $1.06_{-0.6}^{+0.8}$ & 15.7/19 & 1.44 && & & \\ 
13 &(\timeform{11h55m20.3s}, \timeform{23D40'32''}) & $0.1_{-1.0}^{+1.0}$ & $<{2.8}$ & 18.7/18 & 1.44 && & & \\ 
14 &(\timeform{11h55m29.7s}, \timeform{23D47'26''}) & $2.0_{-0.5}^{+0.7}$ & $2.7_{-1.3}^{+1.7}$ & 22.4/14 & 1.28 && & & \\ 
15 &(\timeform{11h54m47.7s}, \timeform{23D39'23''}) & $1.5_{-0.3}^{+0.3}$ & $6.6_{-1.8}^{+2.0}$ & 23.2/19 & 1.43 && & & \\ 
\makebox[0in][l]{Total} & & $\makebox[0in][r]{1}.82_{-0.12}^{+0.12}$ & $\makebox[0in][r]{4}8.3_{-5.6}^{+6.0}$ & 113.1/117 & 1.36 &&			$\makebox[0in][r]{1}.92_{-0.09}^{+0.09}$&$\makebox[0in][r]{3}2.3_{-4.4}^{+4.8}$ & 87.2/77 \\
\hline \\[-1ex]
\multicolumn{10}{l}{\parbox{0.97\textwidth}{\footnotesize\footnotemark[$\ast$]
Source--04, 05, and 08 are excluded because they exhibited negative
counts after the background subtraction.}}\\
\multicolumn{10}{l}{\parbox{0.97\textwidth}{\footnotesize\footnotemark[$\dag$]
Source--11, 12, 13, 14, and 15 are out of MOS1 and MOS2 FOVs.}}\\
\multicolumn{10}{l}{\parbox{0.97\textwidth}{\footnotesize\footnotemark[$\ddag$]
Unit of flux is  10$^{-14}$ \fluxunit\ (2--10 keV).}}\\
\end{tabular}
\end{center}
\end{table*}

\subsection{Cosmic X-ray Background}\label{sec:CXB}

An ICM temperature measurement in the outer regions of a cluster is
very sensitive to the CXB level. We took the 100\% CXB surface
brightness to be $I_0 = 6.38\times10^{-8}$ \sbunit\ based on
the ASCA-GIS measurements \citep{kushino02}. \citet{Moretti2008}
summarized measurements \citep{Gruber1999, McCammon1983, Gendreau1995,
Vecchi1999,kushino02,Revnivtsev2003,DeLuca2004,Revnivtsev2005,Hickox2006}
of the CXB level, including their new result with {\it XMM-Newton}.
The measured CXB surface brightnesses show a significant range from
the HEAO1 value of $(5.41\pm 0.56) \times 10^{-8}$ \sbunit\ \citep{Gruber1999}
to $(7.71\pm 0.33)\times 10^{-8}$ \sbunit\  with SAX-MECS
\citep{Vecchi1999} in the 2--10~keV band.  Recent measurements show
the flux to be within about 10\% of the level reported by
\citet{kushino02}.

We estimated the remaining CXB surface brightness after the above
point-source subtraction by the following three methods: (1)
subtracting the summed point source fluxes measured with Suzaku
from the 100\% CXB, (2) subtracting the summed point source
fluxes estimated using the $\log N$--$\log S$ relation,
and (3) fitting a power-law model to the diffuse
emission in the $20'-26'$ region after the point sources are excised.

In case (1), we subtracted contribution of the excised sources of
$1.80_{-0.21}^{+0.22}\times 10^{-8}$ \sbunit\ from the 100\% CXB,
dividing $F_{\rm X} = 4.83^{+0.60}_{-0.56}\times 10^{-13}$ \fluxunit\ of
the Suzaku total by $17'.8\times 17'.8$ area of the XIS FOV\@.
In case (2), we calculated the
integrated point source flux per steradian from
\begin{eqnarray}\label{eq:fs}
I(S>S_{0}) &=& \frac{k_0}{\gamma -2}\; S_{0}^{-\gamma+2},
\end{eqnarray}
where k and $\gamma$ are the differential $\log N$--$\log S$ normalization
and slope, respectively. We take nominal values,
$k_0=1.58\times 10^{-15}$~sr$^{-1}$~(\fluxunit)$^{\gamma-1}$
and $\gamma=2.5$, from \citet{kushino02}.
$S_{0}$ is taken as $2\times 10^{-14}$ \fluxunit,
which is slightly higher than our flux limit, because the assumed
$\log N$--$\log S$ in equation (\ref{eq:fs}) does not take into account
the flattening of the relation in the fainter flux end.
In case (3), we fit the spectra from the solid angle in
the $20'-26'$ annulus that remain after the source excision by a
power-law model using a uniform flux ancillary response file
(ARF; see section~\ref{sec:resp}). The ARF assumes that
X-ray photons comes into the detectors uniformly from the sky direction
within 20$'$ radius from the optical axes of the respective XRTs.
The model fit is ${\it apec} + {\it wabs}\times ({\it apec} + {\it powerlaw})$
where the two {\it apec} components represent the galactic emission.
This is the 2T-III model described in section \ref{sec:GAL}.
In this case, value of the $I_0 - I_{\rm X}$ column is determined
by the spectral fit, and then $I_{\rm X}$ is calculated assuming
$I_0 = 6.38\times10^{-8}$ \sbunit\ in table~\ref{tab:cxb}.

We summarize our estimations of the remaining CXB surface brightness,
$I_0 - I_{\rm X}$, in table~\ref{tab:cxb}.
All three methods give consistent results. 
Hereafter we will use a nominal diffuse cosmic X-ray background
spectrum (after subtraction of point sources brighter than
$\sim 1\times 10^{-14}$ \fluxunit\ in 2--10 keV band)
described by a power-law with a photon index $\Gamma=1.37$,
and surface brightness $4.73\times 10^{-8}$ \sbunit\ in the 2--10 keV band,
which comes from the 2T-III~(a) row of the method (3).
We adopt this method because it directly measures the quantity of
interest in our observations.

To estimate the amplitude of the CXB fluctuations, we scaled the
measured fluctuations from Ginga \citep{Hayashida1989} to
our flux limit and FOV area. The fluctuation width is given by
the following relation,
\begin{eqnarray}
\frac{\sigma_{\rm Suzaku}}{I_{\rm CXB}} = \frac{\sigma_{\rm Ginga}} 
{I_{\rm CXB}} \left(\frac{\Omega_{\rm e,Suzaku}}{\Omega_{\rm e,Ginga}} 
\right)^{-0.5} \left(\frac{S_{\rm c,Suzaku}}{S_{\rm c,Ginga}} 
\right)^{0.25},
\end{eqnarray}
where $(\sigma_{\rm Suzaku}/I_{\rm CXB})$ means the fractional CXB
fluctuation width due to the statistical fluctuation of discrete
source number in the FOV\@. Here, we adopt
$\sigma_{\rm Ginga}/I_{\rm CXB} = 5\%$,
with $S_{\rm c}$ (Ginga: $6\times10^{-12}$ \fluxunit)
representing the upper cut-off of the source flux,
and $\Omega_{\rm e}$ (Ginga: 1.2~deg$^2$) representing
the effective beam size (or effective solid angle) of the detector.
We show the result, $\sigma$/$I_{\rm CXB}$,
for each spatial region in table~\ref{tab:region}.

\begin{table*}[bt]
\caption{
Estimation of the CXB surface brightness after the point source excision.
}\label{tab:cxb}
\begin{center}
\begin{tabular}{lccc}
\hline\hline
\makebox[31em]{} & $I_0 - I_{\rm X}$ $^\ast$ & $I_{\rm X}$ $^\dag$ & $\Gamma$ $^\ddag$ \\ \hline
(1) $\dotfill$ &				$4.58_{-0.21}^{+0.22}$ & $1.80_{-0.21}^{+0.22}$ & \makebox[0in][c]{1.41 (fixed)}\\
(2) $^{\S}$ $\dotfill$ &			\multicolumn{1}{l}{4.15} & \multicolumn{1}{l}{2.23} & \makebox[0in][c]{1.41 (fixed)}\\
(3) 2T-III (a) $^{\|}$ $\dotfill$ &		$4.73_{-0.22}^{+0.13}$ & $1.65_{-0.22}^{+0.13}$ & $1.37_{-0.05}^{+0.04}$\\
(3) 2T-III (b) $^{\sharp}$ $\dotfill$ &		$4.69_{-0.18}^{+0.18}$ & $1.69_{-0.18}^{+0.18}$ & $1.40_{-0.07}^{+0.05}$\\
(3) 2T-III (c) $^{\|}$ $\dotfill$ &		$5.16_{-0.58}^{+0.12}$ & $1.22_{-0.58}^{+0.12}$ & $1.44_{-0.05}^{+0.03}$\\
(3) contami$+$20\% $^{\|}$ $\dotfill$ &		$5.04_{-0.35}^{+0.16}$ & $1.34_{-0.35}^{+0.16}$ & $1.45_{-0.05}^{+0.05}$\\
(3) contami$-$20\% $^{\|}$ $\dotfill$ &		$4.95_{-0.33}^{+0.13}$ & $1.33_{-0.33}^{+0.13}$ & $1.44_{-0.04}^{+0.06}$\\
\hline \\[-1ex]
\multicolumn{4}{l}{\parbox{0.97\textwidth}{\footnotesize$^{\ast}$
Estimated surface brightness of the CXB after the point source excision
in unit of $10^{-8}$ \sbunit\ (2--10~keV)\@.
}}\\
\multicolumn{4}{l}{\parbox{0.97\textwidth}{\footnotesize$^{\dag}$
Contribution of the resolved point sources
in unit of $10^{-8}$ \sbunit\ (2--10~keV)\@.
}}\\
\multicolumn{4}{l}{\parbox{0.97\textwidth}{\footnotesize$^{\ddag}$
Assumed or estimated photon index of the CXB\@.
}}\\
\multicolumn{4}{l}{\parbox{0.97\textwidth}{\footnotesize$^{\S}$
Surface brightness of 100\% of CXB is assumed as
$I_0 = 6.38\times 10^{-8}$ \sbunit\ (2--10~keV)\@.
Integrated point source contribution, $I_{\rm X}$,
is calculated with equation (\ref{eq:fs}).
See section \ref{sec:CXB} for details.
}}\\
\multicolumn{4}{l}{\parbox{0.97\textwidth}{\footnotesize$^{\|}$  
See section \ref{sec:GAL} for definition.
Abundance model is by \citet{anders89}.
   }}\\
\multicolumn{4}{l}{\parbox{0.97\textwidth}{\footnotesize$^{\sharp}$  
See section \ref{sec:GAL} for definition.
Abundance model is by \citet{Feldman1992}.
   }}\\
\end{tabular}
\end{center}
\end{table*}

\begin{table*}
\caption{
Properties of the spatial regions used.
}\label{tab:region}
\centerline{\small
\begin{tabular}{rrrrrrrrrrr}
\hline\hline
\makebox[3em][l]{Region\,$^\ast$} & \makebox[1.5em]{} & \multicolumn{1}{c}{$\Omega_{\rm e}$ 
\makebox[0in][l]{\,$^\dagger$}} & \multicolumn{1}{c}{\it Coverage\makebox[0in][l]{\,$^\dagger$}} & \makebox[4.2em][r]{\scriptsize\it SOURCE\_ 
\makebox[0in][l]{\,$^\ddagger$}} & \makebox[4em][l]{$ 
\sigma$/$I_{\rm CXB}$ $^{\S}$}\hspace*{-0.5em} &
\multicolumn{5}{c}{FI counts (0.5--10 keV)\makebox[0in][l]{\,$^\|$}} \\
\cline{7-11}
&& \makebox[1em][c]{(arcmin$^2$)} & &\makebox[4.2em][r]{\scriptsize\it
RATIO\_REG} & \makebox[4em][c]{\scriptsize (\%)}& OBS & NXB & CXB &  
GAL & $f_{\rm BGD}$ \\
\hline
$2'.7-7'$ & $\dotfill$
&$\!\!18.6$&$\!\!14.2\%$&$\!\!2.60\%$&$\!\!15.4$\hspace*{1em} &$\!\! 
3,828\pm62$&$\!\!855\pm86$&$\!\!560\pm86$&$\!\!81\pm9$&$\!\! 
39.1\pm3.5\%$\\
$7'-10'$ & $\dotfill$
&$\!\!25.6$&$\!\! 16.0\%$&$\!\!1.49\%$&$\!\!13.1$\hspace*{1em} &$\!\! 
3,568\pm60$&$\!\!1,241\pm124$&$\!\!966\pm127$&$\!\!131\pm11$&$\!\! 
65.5\pm5.3\%$\\
$10'-15'$ & $\dotfill$
&$\!\!55.0$&$\!\!14.0\%$&$\!\!1.40\%$&$\!\!9.0$\hspace*{1em} &$\!\! 
6,340\pm80$&$\!\!2,428\pm242$&$\!\!2,460\pm220$&$\!\!296\pm17$&$\!\! 
81.8\pm5.3\%$\\
$15'-20'$ & $\dotfill$
&$\!\!86.5$&$\!\!15.7\%$&$\!\!1.27\%$&$\!\!7.1$\hspace*{1em} &$\!\! 
9,156\pm96$&$\!\!4,162\pm416$&$\!\!4,035\pm288$&$\!\!499\pm22$&$\!\! 
95.0\pm5.6\%$\\
$20'-26'$ & $\dotfill$
&$\!\!38.6$&$\!\! 4.5\%$&$\!\!0.47\%$&$\!\!10.7$\hspace*{1em} &$\!\! 
4,547\pm67$&$\!\! 2,523\pm252$&$\!\!1,907\pm204$&$\!\!272\pm16$&$\!\! 
103.4\pm7.3\%$\\
\hline\hline
\makebox[3em][l]{Region\,$^\ast$} & & \multicolumn{1}{c}{$\Omega_{\rm e}$ 
\makebox[0in][l]{\,$^\dagger$}} & \multicolumn{1}{c}{\it Coverage\makebox[0in][l]{\,$^\dagger$}} & \makebox[4.2em][r]{\scriptsize\it SOURCE\_ 
\makebox[0in][l]{\,$^\ddagger$}}& \makebox[4em][l]{$\sigma 
$/$I_{\rm CXB}$ $^{\S}$}\hspace*{-0.5em} &
\multicolumn{5}{c}{BI counts (0.4--10.0 keV)\makebox[0in][l]{\,$^\|$}} \\
\cline{7-11}
&& \makebox[1em][c]{(arcmin$^2$)} & &\makebox[4.2em][r]{\scriptsize\it
RATIO\_REG}&\makebox[4em][c]{\scriptsize (\%)}\hspace*{-0.5em}& OBS &  
NXB & CXB & GAL & $f_{\rm BGD}$ \\
\hline
$2'.7-7'$ & $\dotfill$
&$\!\!18.4 $&$\!\!14.0\%$&$\!\!2.60\%$&$\!\!15.5$\hspace*{1em} &$\!\! 
2,042\pm45$&$\!\!748\pm75$&$\!\!208\pm32$&$\!\!85\pm9$&$\!\! 
50.9\pm4.6\%$\\
$7'-10'$ & $\dotfill$
&$\!\! 25.5 $&$\!\!15.9\%$&$\!\!1.56\%$&$\!\!13.1$\hspace*{1em} &$\!\! 
2,546\pm50$&$\!\!1,088\pm109$&$\!\!392\pm51$&$\!\!144\pm12$&$\!\! 
63.8\pm5.1\%$\\
$10'-15'$ & $\dotfill$
&$\!\!54.9 $&$\!\! 14.0\%$&$\!\!1.41\%$&$\!\!9.0$\hspace*{1em} &$\!\! 
4,447\pm67$&$\!\!2,277\pm228$&$\!\!1,012\pm91$&$\!\!331\pm18$&$\!\! 
81.4\pm5.7\%$\\
$15'-20'$ & $\dotfill$
&$\!\!87.4 $&$\!\! 15.9\%$&$\!\!1.32\%$&$\!\!7.1$\hspace*{1em} &$\!\! 
7,425\pm86$&$\!\!4,418\pm441$&$\!\!1,857\pm132$&$\!\!631\pm25$&$\!\! 
93.0\pm6.3\%$\\
$20'-26'$ & $\dotfill$
&$\!\! 24.6$&$\!\! 2.8\%$&$\!\!0.37\%$&$\!\!13.3$\hspace*{1em} &$\!\! 
2,984\pm55$&$\!\!1,954\pm195$&$\!\!706\pm94$&$\!\!264\pm16$&$\!\! 
98.0\pm7.5\%$\\
\hline\\[-1ex]
\multicolumn{11}{l}{\parbox{0.97\textwidth}{\footnotesize
\footnotemark[$\ast$]
Radii are from the XMM-Newton surface brightness peak
in figure~\ref{fig:image}(a).
}} \\
\multicolumn{11}{l}{\parbox{0.97\textwidth}{\footnotesize
\footnotemark[$\dagger$]
The average value of the four detectors.
}} \\
\multicolumn{11}{l}{\parbox{0.97\textwidth}{\footnotesize
\footnotemark[$\ddagger$]
$\makebox{\it SOURCE\_RATIO\_REG}\equiv\makebox{\it Coverage}\;
\times\int_{r_{\rm in}}^{r_{\rm out}} S(r)\; r\,dr /  
\int_{0}^{\infty} S(r)\; r\,dr $,where $S(r)$ represents the assumed  
radial profile of A1413. We confined $S(r)$ to a $60'\times 60'$
region on the sky.
}}\\
\multicolumn{11}{l}{\parbox{0.97\textwidth}{\footnotesize
\footnotemark[$\S$]
$S_{\rm c} = 1\times 10^{-14}$ \fluxunit\  is assumed for all regions.
}}\\
\multicolumn{11}{l}{\parbox{0.97\textwidth}{\footnotesize
\footnotemark[$\|$]
OBS denotes the total observed counts. NXB, CXB and GAL are the estimated 
counts. $f_{\rm BGD}\equiv ({\rm NXB + CXB + GAL})/{\rm OBS}$.
}}
\end{tabular}
}
\end{table*}

\subsection{Non X-ray Background}
The non X-ray background (NXB) spectra were estimated from the Suzaku
database of dark earth observations using the procedure of
\citet{tawa08}. We accumulated data for the same detector area, for
the same distribution of COR2 as the A1413 observation using
the {\it xisnxbgen} FTOOLS covering 30 days before to 90 days after the
observation period of A1413. To increase the A1413 signal-to-noise
ratio by reducing the NXB count rate, we required COR2 to be $> 8$ GV
and PINUD to be between 100 and 300~cts~s$^{-1}$. After this
screening the exposure time dropped from 108 ks to 72 ks, nevertheless
the fit residuals were reduced. We also tested other screening
criteria, such as COR2 $>8$ GV and COR2 $>5$ GV, both with no PINUD
screening. The former criterion did not affect the final spectral
results significantly, but the latter gave different ICM temperatures.
To test a possible NXB uncertainity systematic error, we varied its
intensity by $\pm 3\%$ as investigated by \citet{tawa08}.

\subsection{Galactic Components}\label{sec:GAL}
We fit the data in the $20'-26'$ region to constrain the foreground
Galactic emission, using the same uniform-sky ARF as the CXB component.
We investigated the best model to use and the best-fit
model parameters. In all cases, we also included a power-law model
to represent the CXB\@. We tried a single temperature thermal plasma model,
1T: ${\it apec} + {\it wabs}\times {\it powerlaw}$, a two temperature model,
2T: ${\it wabs}\times ({\it apec}_1 + {\it apec}_2 + powerlaw)$,
and a two temperature model following \citet{tawa09},
2T-III: ${\it apec}_1 + {\it wabs}\times ({\it apec}_2 + powerlaw)$.
In all models, redshift and abundance of the {\it apec}
components were fixed at 0.0 and 1.0, respectively. The two
temperature variants try to model the Local Hot Babble (LHB) and the
Milky Way Halo (MWH).
We tried three types of the 2T model:
both temperatures fixed to 0.204~keV and 0.074~keV given by \citet{Lumb2002},
one temperature fixed to 0.074~keV and the second temperature free,
both temperatures free. We call the first model as 2T-I,
and the second model as 2T-II\@.
The third model did not converge in the fitting,
so that we do not discuss it further.

We found that the 1T and 2T-I models gave worse $\chi^2$ values
compared with the 2T-II and 2T-III fits. 
We show the best-fit parameters in table~\ref{tab:gal} for the 2T-III
model, which we adopt.  We find that the LHB and MWH temperatures are
$0.112_{-0.005}^{+0.009}$ keV and $0.278_{-0.019}^{+0.029}$ keV,
respectively. These values are consistent with those obtained by
\citet{tawa09}. We also show in table~\ref{tab:gal} how the best-fit
parameters change as a result of systematic changes in the CXB and NXB
levels and of the abundance model used (labeled (a) or (b)). The
variations are small: less than $\pm 10\%$ for the temperatures and
$\pm 15\%$ for the normalizations. Finally, our baseline CXB+GAL
model is denoted 2T-III (a),
${\it apec}_1 + {\it wabs}*({\it apec}_2 + powerlaw)$ with
abundances from \citet{anders89}. We link all parameters of this
model, except an overall normalization, when performing the fits for
the different spatial regions described in section \ref{sec:specana}.

\begin{table*}[bt]
\caption{
Galactic components best fit parameters and 90\% confidence errors.
}\label{tab:gal}
\begin{center}
\begin{tabular}{lcccccc}
\hline\hline
\makebox[11em]{} & $kT_{1}$ (keV)& $Norm_{1}$ $^{\S}$&  $S_{1}$ $^{\|}$ & $kT_{2}$ (keV) & $ Norm_{2}$ $^{\S}$ & $S_{2}$ $^{\|}$ \\  
\hline
2T-III (a) $^{*}$ $\dotfill$ & $0.112_{-0.005}^{+0.009}$ &  $1.994_{-0.163}^{+0.147}$ & $0.616_{-0.050}^{+0.046}$ &  $0.278_{-0.019}^{+0.029}$ & $0.194_{-0.037}^{+0.051}$ &  $0.258_{-0.050}^{+0.068}$\\
2T-III (b) $^{\dagger}$  $\dotfill$ & $0.110_{-0.006}^{+0.003}$ &  $2.199_{-0.154}^{+0.148}$ & $0.633_{-0.044}^{+0.043}$ &  $0.314_{-0.025}^{+0.029}$ & $0.222_{-0.043}^{+0.045}$ &  $0.302_{-0.059}^{+0.062}$\\
2T-III (c) $^{\ddagger}$  $\dotfill$ & $0.113_{-0.003}^{+0.003}$ & $1.727_{-0.101}^{+0.181}$ & $0.560_{-0.033}^{+0.059}$ & $0.260_{-0.033}^{+0.964}$ & $0.201_{-0.028}^{+0.065}$ & $0.266_{-0.037}^{+0.086}$\\
NXB+3\%, CXB$_{\rm MAX}$ $\dotfill$ & $0.112_{-0.005}^{+0.003}$ &  $2.015_{-0.105}^{+0.180}$ & $0.640_{-0.033}^{+0.057}$ &  $0.311_{-0.028}^{+1.183}$ & $0.197_{-0.036}^{+0.042}$ &  $0.269_{-0.049}^{+0.057}$\\
NXB$-$3\%, CXB$_{\rm MIN}$ $\dotfill$ & $0.111_{-0.006}^{+0.014}$ &  $2.170_{-0.203}^{+0.239}$ & $0.651_{-0.061}^{+0.072}$ &  $0.319_{-0.026}^{+0.706}$ & $0.227_{-0.074}^{+0.033}$ &  $0.311_{-0.101}^{+0.046}$\\
contami+20\% $\dotfill$ & $0.111_{-0.010}^{+0.002}$ & $2.254_{-0.222}^{+0.127}$ &  $0.660_{-0.065}^{+0.037}$ & $0.269_{-0.013}^{+0.818}$ &  $0.223_{-0.044}^{+0.052}$ & $0.295_{-0.058}^{+0.069}$\\
contami$-$20\% $\dotfill$ & $0.113_{-0.006}^{+0.005}$ & $1.791_{-0.152}^{+0.160}$ &  $0.569_{-0.048}^{+0.051}$ & $0.286_{-0.070}^{+0.920}$ &  $0.173_{-0.039}^{+0.055}$ & $0.232_{-0.053}^{+0.073}$\\
\hline \\[-1ex]
\multicolumn{7}{l}{\parbox{0.97\textwidth}{\footnotesize$^{*}$
Abundance model is by \citet{anders89}.}}\\
\multicolumn{7}{l}{\parbox{0.97\textwidth}{\footnotesize$^{\dagger}$
Abundance model is by \citet{Feldman1992}.}}\\
\multicolumn{7}{l}{\parbox{0.97\textwidth}{\footnotesize$^{\ddagger}$
Including two gaussian models of O$_{\rm VII}$ and O$_{\rm VIII}$.
Abundance model is by \citet{anders89}.}}\\
\multicolumn{7}{l}{\parbox{0.97\textwidth}{\footnotesize$^{\S}$  
Normalization of the apec component scaled with a factor of
${\it SOURCE\_RATIO\_REG} / \Omega_{\rm e}$ in table~\ref{tab:region},\\
${\it Norm} = {\it SOURCE\_RATIO\_REG}/\Omega_{\rm e}\;
\int n_{\rm e} n_{\rm H} dV \,/\, (4\pi\, (1+z)^2 D_{\rm  
A}^{\,2})\times 10^{-20}$~cm$^{-5}$~arcmin$^{-2}$, where $D_{\rm A}$ is  
the angular diameter distance to the source.}}\\
\multicolumn{7}{l}{\parbox{0.97\textwidth}{\footnotesize$^{\|}$
Surface brightness in unit of
10$^{-6}$ photons~cm$^{-2}$~s$^{-1}$~arcmin$^{-2}$ (0.5--10 keV)\@.
}}
\end{tabular}
\end{center}
\end{table*}

\subsection{Background Fraction in Each Region}
Table~\ref{tab:region} presents many properties of the spatial regions
we analyzed. The columns are the annular boundaries; the actual solid
angle of each region observed, $\Omega_{\rm e}$; the coverage fraction of
each annulus which is the ratio of $\Omega_{\rm e}$ to the total solid
angle of the annulus, {\it Coverage}; the fraction of the simulated cluster
photons which fall in the region compared with the total photons from
the entire simulated cluster, {\it SOURCE\_RATIO\_REG};
the CXB fluctuations due to unresolved point sources, $\sigma/I_{\rm CXB}$;
the observed counts, OBS;
the estimated counts for each background component, NXB, CXB, and GAL;
and the fraction of background photons given by
$f_{\rm BGD}\equiv$ (NXB+CXB+GAL)/OBS\@.

The NXB count rates are calculated from the dark earth data. We
simulated the CXB and GAL components spectra using {\it xissim} with
the flux and spectral parameters given in row 2T-III~(a) of tables
\ref{tab:cxb} and \ref{tab:gal}, assuming a uniform surface
brightness that fills the field. We plot the NXB and CXB spectra in
figures \ref{fig:spec-fi} and \ref{fig:spec-bi}. These spectra gave
the count rates in table~\ref{tab:region}. In the outermost region of
$20'-26'$, $f_{\rm BGD}$ is consistent with 100\%. This confirms
the accuracy of our background estimation.

\section{Spectral Analysis}\label{sec:specana}
\subsection{Spatial and spectral responses}\label{sec:resp}
We need to prepare the spatial and spectral responses which are
necessary for reducing and analyzing our observations of A1413. These
responses have complicated properties for extended sources. Indeed
they depend on the surface brightness distribution of the source and
so are unique for each annular region. Monte Carlo simulators are used to
generate some of the responses. The X-ray telescope + XIS
simulator is called {\it xissim}, and the ARF generator using
the simulator is called {\it xissimarfgen} \citep{ishisaki07}.
We used version 2008-04-05 of the simulator.

A surface brightness distribution is necessary for {\it xissim}
and {\it xissimarfgen},
because the point spread function (PSF) of the XRT
produces an efficiency that is correlated among adjacent spatial cells.
Since the XIS FOV did not include the
brightness peak of A1413, we used the KBB model of \citet{pratt02} to
generate the ARF\@. We numerically projected the KBB 3-dimensional
model of the gas density to generate the input surface brightness
distribution. Since the ARF describes the detection efficiency as a
function of energy, no particular spectral shape is required for
input. The effect of the XIS IR/UV blocking filter contamination is
included in the ARF based on the calibration of November 2006. The
normalization of the ARF is such that the measured flux in a spectral
fit for a given spatial region is the flux from the entire input
surface brightness. The flux just from the spatial region is the fit
flux times the {\it xissimarfgen} output parameter {\it SOURCE\_RATIO\_REG}
(table~\ref{tab:region}).
The surface brightness from a given spatial region is then the
usual flux from the region divided by the solid angle that
subtends from the observer.

We examined how many photons accumulated in the five annular regions
actually came from somewhere else on the sky because of the extended
telescope PSF\@. We show in table~\ref{tab:stray} the results for the
FI+BI detectors in the 0.5--5 keV band. These numbers agree well
within 1\% for individual sensors and other reasonable energy
bands. About 70\% of the photons detected in each region actually come
from the corresponding sky region.  \citet{serlemitsos07} gives an
upper limit on the error in the simulation at $20'$. He reported
that the actual stray intensity levels were less than twice those
predicted by {\it xissim} due to the XRT reflector alignment errors and
reflections from the pre-collimator blades.

\begin{table*}[bt]
\caption{
Emission weighted radius and estimated fractions of the ICM
photons accumulated in detector regions coming from each
sky region for FI+BI in the 0.5--5~keV band.
}\label{tab:stray}
\begin{center}
\begin{tabular}{rrrrrrrrr}
\hline\hline
Detector & \makebox[11em]{} & \multicolumn{1}{l}{\makebox[0in][c]{Emission weighted}} \hspace*{2em} & \multicolumn{6}{c} 
{Sky region} \\
\cline{4-9}
region & & \multicolumn{1}{c}{radius $^{\ast}$} & \makebox[2.8em][c]{0--2.7$'$} & 
\makebox[2.8em][c]{2.7--7$'$} & \makebox[2.8em][c]{7--10$'$} &  
\makebox[2.8em][c]{10--15$'$} & \makebox[2.8em][c]{15--20$'$}  
& \makebox[2.8em][c]{20--26$'$} \\
\hline
$2'.7-7'$ & $\dotfill$ & $ 4.7_{-2.0}^{+2.3}$ \hspace*{2em} & 21.5\% & 73.2\%  &  5.1\% &  0.2\% &  0.0\% &  0.0\%\\
$7'-10'$  & $\dotfill$ & $ 8.0_{-1.0}^{+2.0}$ \hspace*{2em} & 16.0\% & 21.8\% &  54.9\% &  7.3\% &  0.1\% &  0.0\%\\
$10'-15'$ & $\dotfill$ & $11.0_{-1.0}^{+4.0}$ \hspace*{2em} &  6.7\% &  7.3\% &  14.0\% & 67.3\% &  4.7\% &  0.0\%\\
$15'-20'$ & $\dotfill$ & $18.6_{-3.6}^{+1.5}$ \hspace*{2em} &  4.1\% &  2.7\% &   2.7\% & 16.8\% & 67.1\% &  5.4\%\\
$20'-26'$ & $\dotfill$ & ------ \hspace*{3em}		    	&  6.7\% &  6.7\% &   0.0\% &  0.0\% &  20.0\% & 66.7\%\\
\hline \\[-1ex]
\multicolumn{9}{l}{\parbox{0.97\textwidth}{\footnotesize$^{\ast}$
Emission weighted radius from the surface brightness peak of the XMM-Newton.}}
\end{tabular}
\end{center}
\end{table*}

The redistribution matrix file (RMF), which gives the spectral
response to a mono-energetic input, is the same for all sources. It
was generated with {\it xisrmfgen} version 2007-05-14.  Degradation of
the energy resolution is included based on the calibration in November
2006.

\subsection{Model Fitting}
\begin{figure*}[htbp]
\begin{minipage}{\textwidth}
\parbox{0.48\textwidth}{\ (a) FI, $2'.7-7'$}
\parbox{0.48\textwidth}{\ (b) FI, $7'-10'$}
\end{minipage}
\begin{minipage}{\textwidth}
\FigureFile(0.48\textwidth,\textwidth){figure3a.eps}
\FigureFile(0.48\textwidth,\textwidth){figure3b.eps}
\medskip
\end{minipage}
\begin{minipage}{\textwidth}
\parbox{0.48\textwidth}{\ (c) FI, $10'-15'$}
\parbox{0.48\textwidth}{\ (d) FI, $15'-20'$}
\end{minipage}
\begin{minipage}{\textwidth}
\FigureFile(0.48\textwidth,\textwidth){figure3c.eps}
\FigureFile(0.48\textwidth,\textwidth){figure3d.eps}
\medskip
\end{minipage}
\begin{minipage}{\textwidth}
\parbox{0.48\textwidth}{\ (e) FI, $20'-26'$}
\end{minipage}
\begin{minipage}{\textwidth}
\FigureFile(0.48\textwidth,\textwidth){figure3e.eps}
\end{minipage}
\caption{
Spectra for the individual annular regions observed with the FI sensors.
The total observed spectrum minus the estimated NXB is the black 
crosses, the estimated NXB is the grey crosses, and the fitted CXB 
component is the solid line. The screening used are COR2 $> 8$ GV and 
100 $<$ PINUD $<$ 300~cts~s$^{-1}$.
$^{55}$Fe calibration source regions, namely {\it calmask},
are excluded except for (a).
}\label{fig:spec-fi}
\end{figure*}

\begin{figure*}[htbp]
\begin{minipage}{\textwidth}
\parbox{0.48\textwidth}{\ (a) BI, $2'.7-7'$}
\parbox{0.48\textwidth}{\ (b) BI, $7'-10'$}
\end{minipage}
\begin{minipage}{\textwidth}
\FigureFile(0.48\textwidth,\textwidth){figure4a.eps}
\FigureFile(0.48\textwidth,\textwidth){figure4b.eps}
\medskip
\end{minipage}
\begin{minipage}{\textwidth}
\parbox{0.48\textwidth}{\ (c) BI, $10'-15'$}
\parbox{0.48\textwidth}{\ (d) BI, $15'-20'$}
\end{minipage}
\begin{minipage}{\textwidth}
\FigureFile(0.48\textwidth,\textwidth){figure4c.eps}
\FigureFile(0.48\textwidth,\textwidth){figure4d.eps}
\medskip
\end{minipage}
\begin{minipage}{\textwidth}
\parbox{0.48\textwidth}{\ (e) BI, $20'-26'$}
\end{minipage}
\begin{minipage}{\textwidth}
\FigureFile(0.48\textwidth,\textwidth){figure4e.eps}
\end{minipage}
\vspace*{-2ex}
\caption{
Same as figure~\ref{fig:spec-fi}, but for the BI detector.
All the $^{55}$Fe calibration source regions are excluded.
}\label{fig:spec-bi}
\end{figure*}

We used XSPEC version 12.4.0y for all spectral fitting. The FI and BI
spectra were fitted simultaneously. We employed a
${\it wabs}\times {\it apec}$\/ model for the ICM emission of the cluster.
The {\it wabs} component models
the photoelectric absorption by the Milky Way, parameterized by the
hydrogen column density that was fixed at the 21~cm value \citep{dickey90}.
The {\it apec} is a thermal plasma model.
Its fitting parameters are normalization,
$kT$ and the ICM abundance. The redshift was fixed at the optical
spectroscopic value ($z=0.1427$). Additional fitting parameters are the two
normalizations and temperatures of the GAL components, and the
normalization and photon index of the power-law model for the CXB
component, as described previously. We did not fit the ICM component
in the outermost $20'-26'$ region because we can explain the observed
spectrum without it, as we show in figure~\ref{fig:2t3}(e).
This situation was planned as we wanted to have
an in-field measurement of the background.
In figures~\ref{fig:spec-fi} and \ref{fig:spec-bi},
we compare the intensities of the observed spectra minus
the NXB to the spectra of the NXB and CXB components.
Figure~\ref{fig:spec-fi}(a) shows very strong
Mn-K$_{\alpha}$ line at 5.9~keV from the $^{55}$Fe calibration source,
therefore we ignored the 5--7~keV energy band when we fit the FI spectrum
of this annulus.

\subsection{Results}
\begin{table*}[htbp]
\caption{
Best fitting parameters of the spectral fits with 90\%  
confidence errors for one parameter.
}\label{tab:bestfit}
\begin{center}
\begin{tabular}{rcccrcr}
\hline\hline
\multicolumn{2}{l}{2T-III (a)\makebox[0in][l]{ $^{\ast}$}} & $kT$ & Abundance & $Norm$ $^{\S}$ & $S$ $^{\|}$ & \multicolumn{1}{c}{$\chi^2$/dof} \\
  & \makebox[14em]{} & (keV) & ($Z_{\odot}$)&  & &\\ \hline
$2'.7-7'$ & $\dotfill$ & $7.03_{-1.11}^{+1.57}$ & $0.44_{-0.39}^{+0.62}$ & 
$16.35_{-1.26}^{+1.16}$ & $5.77_{-0.45}^{+0.41}$ & 77.4/107 \\
$7'-10'$ & $\dotfill$ & $4.13_{-0.65}^{+0.97}$ & $0.54_{-0.26}^{+0.21}$ & 
$4.53_{-0.46}^{+0.30}$ & $2.12_{-0.22}^{+0.14}$ & 98.7/116\\
$10'-15'$ & $\dotfill$ & $3.60_{-0.62}^{+0.77}$ & $0.39_{-0.24}^{+0.17}$ & 
$2.29_{-0.25}^{+0.19}$ & $0.90_{-0.10}^{+0.08}$ & 130.1/118\\
$15'-20'$ & $\dotfill$ & $\uparrow$ & $\uparrow$ & 
$0.82_{-0.26}^{+0.11}$ & $0.31_{-0.10}^{+0.04}$ & 109.5/116\\
$20'-26'$ & $\dotfill$ & --- & --- & \multicolumn{1}{c}{---} & --- & 152.7/113 \\
Total & $\dotfill$ & --- & --- & \multicolumn{1}{c}{---} & ---  & 568.4/570\\
\hline\hline
\multicolumn{2}{l}{2T-III (b)\makebox[0in][l]{ $^{\dagger}$}} & $kT$ & Abundance & $Norm$ $^{\S}$ & $S$ $^{\|}$ & \multicolumn{1}{c}{$\chi^2$/dof} \\
  && (keV) & ($Z_{\odot}$)&  & &\\ \hline
$2'.7-7'$ & $\dotfill$ & $7.14_{-1.17}^{+1.62}$ & $0.58_{-0.40}^{+0.42}$ & 
$16.04_{-0.97}^{+2.54}$ & $5.75_{-0.35}^{+0.91}$ & 77.1/107\\
$7'-10'$ & $\dotfill$ & $4.41_{-0.79}^{+0.95}$ & $0.66_{-0.36}^{+0.23}$ & 
$4.43_{-0.46}^{+0.24}$ & $2.11_{-0.22}^{+0.11}$ & 100.6/116\\
$10'-15'$ & $\dotfill$ & $4.03_{-0.66}^{+0.91}$ & $0.77_{-0.51}^{+0.20}$ & 
$2.07_{-0.17}^{+0.12}$ & $0.90_{-0.07}^{+0.05}$ & 129.6/118\\
$15'-20'$ & $\dotfill$ & $\uparrow$ & $\uparrow$ &
$0.72_{-0.23}^{+0.09}$ & $0.31_{-0.10}^{+0.04}$ & 114.7/116\\
$20'-26'$ & $\dotfill$ & --- & --- & \multicolumn{1}{c}{---} & --- & 149.2/113 \\
Total & $\dotfill$ & --- & --- & \multicolumn{1}{c}{---} & ---  & 571.3/570\\
\hline\hline
\multicolumn{2}{l}{2T-III (c)\makebox[0in][l]{ $^{\ddagger}$}} & $kT$ & Abundance & $Norm$ $^{\S}$ & $S$ $^{\|}$ & \multicolumn{1}{c}{$\chi^2$/dof} \\
  && (keV) & ($Z_{\odot}$)&  & &\\ \hline
$2.7'-7'$& $\dotfill$ & 7.20$_{-1.20}^{+1.58}$ & 0.43$_{-0.21}^{+0.22}$ &
26.54$_{-0.90}^{+0.92}$ & 9.41$_{-0.32}^{+0.33}$ & 76.7/105 \\
$7'-10'$& $\dotfill$ & 4.33$_{-0.70}^{+0.92}$ & 0.68$_{-0.22}^{+0.21}$ &
11.02$_{-0.60}^{+0.66}$ & 5.44$_{-0.30}^{+0.33}$ & 99.8/114  \\
$10'-15'$& $\dotfill$ & 3.97$_{-0.66}^{+0.82}$ & 0.53$_{-0.21}^{+0.22}$ &
2.07$_{-0.13}^{+0.14}$ & 0.89$_{-0.06}^{+0.06}$ & 125.2/116 \\
$15'-20'$& $\dotfill$ & $\uparrow$ & $\uparrow$ &
0.66$_{-0.13}^{+0.11}$ & 0.34$_{-0.07}^{+0.06}$ & 104.3/114 \\
$20'-26'$& $\dotfill$ & --- & --- & \multicolumn{1}{c}{---} & --- & 154.5/113\\
Total & $\dotfill$ & --- & --- & \multicolumn{1}{c}{---} & ---  & 560.5/562\\
\hline \\[-1ex]
\multicolumn{7}{l}{\parbox{0.97\textwidth}{\footnotesize$^{\ast}$  
Abundance model is \citet{anders89}.
}}\\
\multicolumn{7}{l}{\parbox{0.97\textwidth}{\footnotesize$^{\dagger}$  
Abundance model is \citet{Feldman1992}.
}}\\
\multicolumn{7}{l}{\parbox{0.97\textwidth}{\footnotesize$^{\ddagger}$  
Including two gaussian models of O$_{\rm VII}$ and O$_{\rm VIII}$ WHIM emission.
Abundance model is \citet{anders89}
}}\\
\multicolumn{7}{l}{\parbox{0.97\textwidth}{\footnotesize$^{\S}$  
Normalization of the apec component scaled with a factor of
${\it SOURCE\_RATIO\_REG}/\Omega_{\rm e}$ in table~\ref{tab:region},\\
${\it Norm} = {\it SOURCE\_RATIO\_REG}/\Omega_{\rm e}\;
\int n_{\rm e} n_{\rm H} dV \,/\, (4\pi\, (1+z)^2 D_{\rm  
A}^{\,2})\times 10^{-20}$~cm$^{-5}$~arcmin$^{-2}$, where $D_{\rm A}$ is  
the angular diameter distance to the source.
}}\\
\multicolumn{7}{l}{\parbox{0.97\textwidth}{\footnotesize$^{\|}$
Surface brightness in unit of
10$^{-6}$ photons~cm$^{-2}$~s$^{-1}$~arcmin$^{-2}$ (0.4--10 keV)\@.
}}
\end{tabular}
\end{center}
\end{table*}

\begin{table*}[htbp]
\caption{
Same as table~\ref{tab:bestfit} except NXB$\pm 3$\%,
CXB$_{\rm MAX}$ and CXB$_{\rm MIN}$ and contami$\pm 20$\%.
Abundance model is \citet{anders89}.
}\label{tab:bestfit-err}
\begin{center}
\begin{tabular}{rcccrcr}
\hline\hline
\multicolumn{2}{l}{NXB$-$3\%, CXB$_{\rm MIN}$} & $kT$ & Abundance & $Norm$ $^{\ast}$ & $S$ $^{\dagger}$ & \multicolumn{1}{c}{$\chi^2$/dof} \\
  & \makebox[14em]{} & (keV) & ($Z_{\odot}$)&  & &\\ \hline
$2'.7-7'$ & $\dotfill$ & $7.57_{-1.28}^{+1.78}$ & $0.47_{-0.31}^{+0.76}$ & 
$16.94_{-1.00}^{+0.70}$ & $5.99_{-0.35}^{+0.25}$ & 78.7/107\\
$7'-10'$ & $\dotfill$ & $4.84_{-0.81}^{+1.11}$ & $0.60_{-0.29}^{+0.32}$ & 
$4.91_{-0.50}^{+0.23}$ & $2.34_{-0.24}^{+0.11}$ & 98.6/116\\
$10'-15'$ & $\dotfill$ & $4.64_{-0.71}^{+0.88}$ & $0.51_{-0.30}^{+0.22}$ & 
$1.07_{-0.18}^{+0.10}$ & $0.43_{-0.07}^{+0.04}$ & 130.6/116\\
$15'-20'$ & $\dotfill$ & $\uparrow$ & $\uparrow$ &
$0.98_{-0.18}^{+0.10}$ & $0.41_{-0.08}^{+0.04}$ & 114.2/116\\
$20'-26'$ & $\dotfill$ & --- & --- & \multicolumn{1}{c}{---} & --- & 157.1/115 \\
Total & $\dotfill$ & --- & --- & \multicolumn{1}{c}{---} & ---  & 579.1/572\\
\hline\hline
\multicolumn{2}{l}{NXB$+$3\%, CXB$_{\rm MAX}$} & $kT$ & Abundance & $Norm$ $^{\ast}$ & $S$ $^{\dagger}$ & \multicolumn{1}{c}{$\chi^2$/dof} \\
  && (keV) & ($Z_{\odot}$)&  & &\\ \hline
$2'.7-7'$ & $\dotfill$ & $6.60_{-1.08}^{+1.57}$ & $0.40_{-0.40}^{+0.86}$ & 
$15.96_{-1.69}^{+0.86}$ & $5.51_{-0.59}^{+0.30}$ & 76.6/107\\
$7'-10'$ & $\dotfill$ & $3.59_{-0.64}^{+0.80}$ & $0.53_{-0.25}^{+0.27}$ & 
$4.16_{-0.72}^{+0.28}$ &$1.87_{-0.32}^{+0.13}$ & 104.3/116\\
$10'-15'$ & $\dotfill$ & $2.52_{-0.39}^{+0.53}$ & $0.35_{-0.19}^{+0.14}$ & 
$2.14_{-0.31}^{+0.18}$ &$0.76_{-0.11}^{+0.06}$ & 130.3/116\\
$15'-20'$ & $\dotfill$ & $\uparrow$ & $\uparrow$ &
$0.53_{-0.19}^{+0.10}$ &$0.18_{-0.06}^{+0.04}$ & 118.6/116 \\
$20'-26'$ & $\dotfill$ & --- & --- & \multicolumn{1}{c}{---} & --- & 150.1/115\\
Total & $\dotfill$ & --- & --- & \multicolumn{1}{c}{---} & ---  & 579.9/572\\
\hline\hline
\multicolumn{2}{l}{contami+20\%} & $kT$ & Abundance & $Norm$ $^{\ast}$ & $S$ $^{\dagger}$ & \multicolumn{1}{c}{$\chi^2$/dof} \\
  && (keV) & ($Z_{\odot}$)&  & &\\ \hline
$2'.7-7'$ & $\dotfill$ &$6.89_{-1.05}^{+1.63}$ &$0.45_{-0.40}^{+0.60}$ & 
$16.31_{-1.20}^{+1.19}$ &$ 5.74_{-0.42}^{+0.42}$ & 77.7/107 \\
$7'-10'$ & $\dotfill$ &$4.01_{-0.63}^{+0.93}$ &$ 0.54_{-0.25}^{+0.25}$ & 
$4.54_{-0.45}^{+0.32}$ &$ 2.10_{-0.21}^{+0.15}$ & 99.0/116 \\
$10'-15'$ & $\dotfill$ &$ 3.17_{-0.51}^{+0.81}$ &$ 0.29_{-0.17}^{+0.17}$ &$  
2.41_{-0.26}^{+0.22}$ &$ 0.90_{-0.10}^{+0.08}$ & 131.3/118 \\
$15'-20'$ & $\dotfill$ & $\uparrow$ & $\uparrow$ &
$0.84_{-0.24}^{+0.13}$ &$ 0.30_{-0.09}^{+0.05}$ & 109.6/116 \\
$20'-26'$ & $\dotfill$ & --- & --- & \multicolumn{1}{c}{---} & --- & 153.4/113 \\
Total & $\dotfill$ & --- & --- & \multicolumn{1}{c}{---} & ---  & 571.0/570\\
\hline\hline
\multicolumn{2}{l}{contami$-$20\%} & $kT$ & Abundance & $Norm$ $^{\ast}$ & $S$ $^{\dagger}$ & \multicolumn{1}{c}{$\chi^2$/dof} \\
  && (keV) & ($Z_{\odot}$)&  & &\\ \hline
$2'.7-7'$ & $\dotfill$ &$7.08_{-1.13}^{+1.56}$ &$ 0.42_{-0.30}^{+0.59}$ & 
$16.36_{-1.08}^{+0.86}$ &$ 5.78_{-0.38}^{+0.30}$ & 77.4/107\\
$7'-10'$ & $\dotfill$ &$ 4.19_{-0.65}^{+0.97}$ &$ 0.54_{-0.25}^{+0.26}$ & 
$4.49_{-0.45}^{+0.31}$ &$ 2.11_{-0.21}^{+0.15}$ & 99.0/116 \\
$10'-15'$ & $\dotfill$ &$ 3.82_{-0.67}^{+0.77}$ &$ 0.44_{-0.26}^{+0.16}$ &$  
2.21_{-0.21}^{+0.17}$ &$ 0.89_{-0.08}^{+0.07}$ & 128.4/118\\
$15'-20'$ & $\dotfill$ & $\uparrow$ & $\uparrow$ &
$0.79_{-0.23}^{+0.10}$ &$ 0.31_{-0.09}^{+0.04}$ & 109.1/116\\
$20'-26'$ & $\dotfill$ & --- & --- & \multicolumn{1}{c}{---} & --- & 153.2/113 \\
Total & $\dotfill$ & --- & --- & \multicolumn{1}{c}{---} & ---  & 567.1/570\\
\hline \\[-1ex]
\multicolumn{7}{l}{\parbox{0.97\textwidth}{\footnotesize$^{\ast}$  
Normalization of the apec component scaled with a factor of
${\it SOURCE\_RATIO\_REG} / \Omega_{\rm e}$ in table~\ref{tab:region},\\
${\it Norm} = {\it SOURCE\_RATIO\_REG}/\Omega_{\rm e}\;
\int n_{\rm e} n_{\rm H} dV \,/\, (4\pi\, (1+z)^2 D_{\rm  
A}^{\,2})\times 10^{-20}$~cm$^{-5}$~arcmin$^{-2}$, where $D_{\rm A}$ is  
the angular diameter distance to the source.
}}\\
\multicolumn{7}{l}{\parbox{0.97\textwidth}{\footnotesize$^{\dagger}$
Surface brightness in unit of
10$^{-6}$ photons~cm$^{-2}$~s$^{-1}$~arcmin$^{-2}$ (0.4--10~keV)\@.
}}
\end{tabular}
\end{center}
\end{table*}

\begin{figure*}[htbp]
\begin{minipage}{\textwidth}
\parbox{0.48\textwidth}{(a) FI+BI, $2'.7-7'$}\hfill
\parbox{0.48\textwidth}{(b) FI+BI, $7'-10'$}
\end{minipage}
\begin{minipage}{\textwidth}
\FigureFile(0.48\textwidth,\textwidth){figure5a.eps}\hfill
\FigureFile(0.48\textwidth,\textwidth){figure5b.eps}
\medskip
\end{minipage}
\begin{minipage}{\textwidth}
\parbox{0.48\textwidth}{(c) FI+BI, $10'-15'$}\hfill
\parbox{0.48\textwidth}{(d) FI+BI, $15'-20'$}
\end{minipage}
\begin{minipage}{\textwidth}
\FigureFile(0.48\textwidth,\textwidth){figure5c.eps}\hfill
\FigureFile(0.48\textwidth,\textwidth){figure5d.eps}
\medskip
\end{minipage}
\begin{minipage}{\textwidth}
\parbox{0.48\textwidth}{(e) FI+BI, $20'-26'$}
\end{minipage}
\begin{minipage}{\textwidth}
\FigureFile(0.48\textwidth,\textwidth){figure5e.eps}
\end{minipage}
\vspace*{-2ex}
\caption{
The upper panels show the observed spectra after subtracting the NXB,
that is fitted with the
ICM: ${\it wabs}\times {\it apec}$ model plus the
GAL+CXB: ${\it apec}_1 + {\it wabs}\times ({\it apec}_2+{\it powerlaw})$
model in the energy range 0.5--10~keV for FI and 0.4--10~keV for BI\@.
The annular regions are: (a) $2'.7-7'$, (b) $5'-10'$, (c) $10'-15'$, 
(d) $15'-20'$, and (e) $20'-26'$. The symbols denote BI data (red 
crosses), FI data (black crosses), CXB of BI (purple),
${\it apec}_1$ of BI (grey),
${\it wabs}\times {\it apec}_2$ of BI (light blue),
ICM of BI (orange),
the total model spectra of BI (green),
and that of FI (blue).
The lower panels show the residuals in units of $\sigma$.
}\label{fig:2t3}
\end{figure*}

\begin{figure*}[htbp]
\begin{minipage}{\textwidth}
\parbox{0.48\textwidth}{\ (a)}\hfill
\parbox{0.48\textwidth}{\ (b)}
\vspace*{-3ex}
\end{minipage}
\begin{minipage}{\textwidth}
\FigureFile(0.48\textwidth,\textwidth){figure6a.eps}\hfill
\FigureFile(0.48\textwidth,\textwidth){figure6b.eps}
\medskip
\end{minipage}
\begin{minipage}{\textwidth}
\parbox{0.48\textwidth}{\ (c)}\hfill
\parbox{0.48\textwidth}{\ (d)}
\vspace*{-3ex}
\end{minipage}
\begin{minipage}{\textwidth}
\FigureFile(0.48\textwidth,\textwidth){figure6c.eps}\hfill
\FigureFile(0.48\textwidth,\textwidth){figure6d.eps}
\end{minipage}
\caption{
Radial profiles for (a) temperature, (b) surface brightness (0.4--10~keV),
(c) abundance, and (d) 3-dimensional electron density. Red diamonds show our Suzaku
results assuming the metal abundances of \citet{anders89}. Orange line
indicates the best-fit profile using the \citet{Feldman1992}
abundances.  Chandra results by \citet{vikhlinin05} are the
black crosses, and the cyan crosses are the XMM-Newton results by 
\citet{snowden08}.  
The uncertainty range due to the combined $\pm 3\%$ variation of the NXB 
level and the maximum/minimum fluctuation of CXB is shown by two blue 
dashed lines. We show by magenta dashed lines the uncertainties 
induced by a $\pm 20\%$ uncertainty in the amount of contamination in 
the IR/UV blocking filters. We also show in panel (b) the CXB level 
(horizontal dashed line) and the Galactic emission (horizontal solid 
line).
}\label{fig:r-profile}
\end{figure*}

In figure~\ref{fig:2t3}, we show the best-fit spectra in each spatial
region. These figures show the observed spectra after subtraction of the NXB,
as well as the best-fit.
These figures show that individual spectra are well fitted by the
model in each region. The normalization for the ICM component was
fixed to zero in the $20'-26'$ annulus to estimate the
background. The ICM spectra did not show strong emission
lines. Because of the low S/N ratio, it was difficult to constrain
the model parameters in the $15'-20'$ annulus.  Therefore, we linked
the ICM temperature and abundance in this region to that of the region
next interior to it, the $10'-15'$ annulus. The best-fit
parameters were consistent within the systematic errors for the two
regions. The emission weighted average radius for the combined region
is $12'.42^{+1'.04}_{-1'.07}$.

Table~\ref{tab:bestfit} shows the best-fit parameters for the ICM
model in each region. We fitted with two different solar abundances,
namely \citet{anders89} and \citet{Feldman1992}. The derived abundance
values are higher when we adopt the \citet{Feldman1992} abundance,
than the \citet{anders89} case, because the Fe abundance relative to H
in the former model is lower than the latter.

In figure~\ref{fig:r-profile}(a), we show temperature profiles
observed with Chandra \citep{vikhlinin05},
XMM-Newton \citep{snowden08}, and Suzaku (this work).
These profiles are consistent with each other in the range $7'-15'$.
The Chandra temperatures are about 20\% higher than the XMM-Newton
values at $2'.7-7'$. The tendency that Chandra gives higher
temperature than XMM-Newton typically becoming significant above
$kT\sim 5$--6~keV is pointed out in figure~12 of \citet{snowden08}.
This discrepancy is due mainly to a Chandra calibration problem,
namely the ground calibration of the HRMA effective area had some errors
especially at the Ir edge (0.62~keV), and there also was uncertainity
about the IR/UV blocking filter contamination. These uncertainties caused a
large discrepancy between the Chandra and XMM-Newton measurements for
high-temperature clusters. Recent updates of the Chandra CALDB,
HRMA AXEFFA version N0008,\footnote{
http://cxc.harvard.edu/ciao4.1/why/caldb4.1.1\_hrma.html}
corrected most of this discrepancy. However, there still remains some
differences in cluster temperature by about 10\% especially in hot
objects.  For temperatures below $\sim 5$~keV, Chandra and XMM-Newton results
are mostly consistent with each other.

We therefore used the XMM-Newton temperatures measured by
\citet{snowden08}.  In fact, their values are higher than those of
\citet{pratt02} who used the same data set.  This difference may
partly be due to the different backgrounds used. Therefore, we
assigned rather large errors of 10\% even in the inner region of $r <
2'.7$ for these data. We will quantify the systematic error of the
Suzaku ICM temperature in the following section.

We plot the related quantities, surface brightness, $S_{\rm X}$,
and 3-dimensional electron density, $n_{\rm e}$,
in figures~\ref{fig:r-profile}(b) and (d). We derived the Chandra
surface brightness from the emission measures provided by A.~Vikhlinin
(private communication). The XMM-Newton surface brightness
is from \citet{snowden08}. The Suzaku surface brightness comes from
the normalization of the {\it apec}\/ model fit. The surface brightness
results are consistent with each other within $10'$. In the outer
region, the Suzaku surface brightness is significantly higher than the
Chandra values.  The cause of this discrepancy could be the different
region of the cluster observed. In particular, Suzaku observed mainly along
the major axis, while Chandra observed the minor axis, as we show in
figure~\ref{fig:image}(a). We obtained the electron density by deprojecting
the emission measure with method describe in \citet{kriss83}.

We show the abundance profile in figure~\ref{fig:r-profile}(c).
Our nominal values are higher than the results of Chandra and
XMM-Newton. However, our errors are large and it is difficult to
draw firm conclusions.

\subsection{Systematic Errors}
To estimate the systematic errors on our electron density, temperature and
abundance profiles, we examined the effects of varying the background
spectra from their nominal levels. We adopted a systematic error for
the NXB intensity of $\pm 3\%$ and the level of the CXB fluctuation
was scaled from the Ginga result \citep{Hayashida1989} as
shown in table~\ref{tab:region}. We considered a $\pm 20\%$ error for
the contamination thickness on the IR/UV blocking filters in front of
the XIS sensors. As mentioned earlier, we also looked into the effect
of the difference between the \citet{anders89} and \citet{Feldman1992}
abundance models.

We give the outcome of these variations in figure~\ref{fig:r-profile}
and table~\ref{tab:bestfit} for the abundance model comparison, and
in figure~\ref{fig:r-profile} and table~\ref{tab:bestfit-err}
for the other comparisons.
Systematic variations of the surface brightness are comparable to
its statistical error for all the systematics we examined.
The same is true of the
temperature except for uncertainties on the UV/IR filter
contamination, where the maximum possible range allowed is about 40\%
larger than the nominal statistical errors. Systematics on the
abundance profile were less than the statistical uncertainties except
for the outer two spatial bins with the \citet{Feldman1992} abundance
models. We conclude from this investigation that our statistical
errors also encompass most possible systematic effects.

\subsection{Search for WHIM lines}
We searched for the warm-hot intergalactic medium (WHIM) which could
exist in the filaments of large-scale structures of the universe. The
outer regions of clusters may be connected to these filaments and are
considered to be promising regions to search for possible WHIM
emission.

We analyzed the regions $2.'7-7'$, $7'-10'$, $10'-15'$, and $15'-20'$.
We fitted the FI+BI spectra simultaneously. We added two gaussian lines to
model the oxygen emission lines. They had fixed redshifted energies of
0.508 keV (O$_{\rm  VII}$) and 0.569 keV (O$_{\rm VIII}$),
with a fixed width of $\sigma=0.0$.
The ICM spectra fitted with the additional two gaussian
lines are shown in figure~\ref{fig:whim}, and table~\ref{tab:bestfit}(c)
gives the fit results. The best temperatures are
consistent with the results of the previous fit without the lines.
Because redshifted line energies overlapped with those of the
Galactic lines, we were unable to distinguish these emission lines
directly. Table~\ref{tab:oxgen} gives our result for the line
intensities which are either $2\sigma$ upper limits or marginal
detections.
\begin{figure*}
\begin{minipage}{\textwidth}
\parbox{0.48\textwidth}{(a) FI+BI, $10'-15'$}\hfill
\parbox{0.48\textwidth}{(b) FI+BI, $15'-20'$}
\end{minipage}
\begin{minipage}{\textwidth}
\FigureFile(0.48\textwidth,\textwidth){figure7a.eps}\hfill
\FigureFile(0.48\textwidth,\textwidth){figure7b.eps}
\end{minipage}
\caption{
O$_{\rm VII}$ (cyan) and O$_{\rm VIII}$ (pink) line spectra
in $10'-15'$ and $15'-20'$ annuli.
}\label{fig:whim}
\end{figure*}

\begin{table}
\caption{
Intensity of redshifted O$_{\rm VII}$ (0.508~keV)
and O$_{\rm VIII}$ (0.569~keV) lines 
in unit of 10$^{-6}$ photons~cm$^{-2}$~s$^{-1}$~arcmin$^{-2}$
with 2$\sigma$ upper limits or 90\% confidence errors
for a single parameter.
}\label{tab:oxgen}
\begin{center}
\begin{tabular}{rccccc}
\hline\hline
\multicolumn{1}{c}{Region} & \makebox[5.5em]{} & $S_{\rm O VIII}$ & $S_{\rm O VII}$ \\ \hline
$2'.7-7'$ & $\dotfill$ & $<0.119$ & $<0.135$\\
$ 7'-10'$ & $\dotfill$ & $<0.075$ & $<0.091$\\
$10'-15'$ & $\dotfill$ & $<0.085$ & $0.094_{-0.061}^{+0.059}$\\
$15'-20'$ & $\dotfill$ & $<0.095$ & $0.081_{-0.051}^{+0.048}$\\
\hline
\end{tabular}
\end{center}
\end{table}

\section{Discussion}
\subsection{Temperature and brightness profiles}

Numerical simulations indicate that the intracluster gas is almost in
hydrostatic equilibrium within the virial radius. For example,
\citet{roncarelli06} showed that the radial density profiles are
smooth out to $\sim 2r_{200}$, while the electron temperature profile
has a discontinuity around 1.3--1.5 $r_{200}$.  \citet{eke98}
performed hydrodynamic simulations in a $\Lambda$CDM universe, and
discussed the possibility of nonequilibrium around $r_{100}$ because
the ratio of kinetic to thermal energy gradually increased from the
center to this radius.

Recent X-ray studies of the outer regions of clusters of galaxies with
Chandra and XMM-Newton showed significant negative temperature
gradients out to a typical radius of $r_{500}$ which is about half of
$r_{200}$ \citep{vikhlinin06, pratt02, snowden08}.
Even though the errors are large, it is significant that our
temperatures continue this steady decline, going from about 7.5 keV
near the center to $\sim 3.5$ keV at $r_{200}$. 
Recent Suzaku results for the A2204 \citep{reiprich09}, PKS0745$-$191
\citep{george08}, and A1795 \citep{bautz09} clusters also show a
temperature drop to 2--3 keV at $r_{200}$. The similar temperatures at
$r_{200}$ are at least partly due to the fact that all these clusters
have similar average temperatures of 5--7 keV\@. What is likely more
significant is the factor of $\sim 2$ decrease in all cases.

We attempted to compare our measured temperature and surface
brightness profiles with theoretical predictions for relaxed
clusters. \citet{suto98} gave ICM properties for clusters whose
potentials follow NFW \citep{navarro96} and modified NFW models,
assuming that the ICM can be described by a polytrope. These models
have 6 parameters and give a wide range of temperature and density
distributions with radius. 

We found that, although we could fit either one of the temperature or
surface brightness profile with the model, it was not possible to fit
both profiles simultaneously despite an exhaustive search of the
6-parameter space. When we fixed the scale radius to be
$r_{s}=350~kpc$ and jointly fit the temperature and brightness
profiles, we obtained reduced $\chi^2$ values of 2.0 using only the
Chandra data and 3.7 for combined Chandra and Suzaku data,
respectively. The likely reason for this result is that the ICM is out
of equilibrium in the outer regions of the cluster.  We examine this
hypothesis in the next section using the entropy profile.

\subsection{Entropy profile}

Entropy carries information about the thermal history of the ICM,
which is thought to be heated by accretion shocks outside the virial
radius.  The central regions of clusters often exhibit complicated
physical phenomena, such as AGN heating and cooling flows, therefore
it is difficult to trace the long-term evolution of clusters there.
In contrast, the outer regions of clusters is where signatures of the
structure formation history can be more clearly seen with the entropy
profiles.  We use the customary X-ray astronomy definition of entropy
as
\begin{eqnarray}
S &=& kT n_e^{-2/3}.
\end{eqnarray}\label{eq:entropy}
We show the entropy profile derived from our data in
figure~\ref{fig:discussion1}(a).  To compare the observed profile with
simulation results, we fit the XMM-Newton data from $0'.5$ to $7'$ and the
Suzaku data from $7'$ to $20'$ with a power-law model,
given by $S \propto r^{\gamma}$. The XMM-Newton data outside of $7'$ have
poorer quality than the Suzaku data, and one Suzaku point inside of
$7'$ was also excluded because it is near the field edge with rather
low data quality.

We found the best-fit power-law indices to be $0.90\pm 0.10$ in
$2'$ to $7'$ and $0.97\pm 0.48$ in $7'$ to $20'$.
The dividing radius of $7'$ corresponds to $0.47\; r_{200}$.
If we fit all the 7 data points
from $2'$ to $20'$, then the slope becomes $0.90\pm 0.12$. These
results indicate that there is no difference in the entropy slopes
between the inner and outer regions.

\citet{voit05} reported $S\propto r^{1.1}$ based on numerical
simulations of adiabatic cool gas accretion, and our observational
result shows a significantly flatter slope, at least for $r < 7'$.
This feature is similar but less pronounced to those reported for
A1795 \citep{bautz09} in which the power-law index flattened ($\gamma
\approx 0.74$) for $r > 4'\sim 0.15\; r_{200}$ and for PKS0745$-$191 where
\citet{george08} also found a flatter entropy profile in the outer
regions. Our result for A1413 suggests that the entropy profile starts
to flatten from $\sim 0.2\; r_{200}$.
To compare the entropy profiles with the simulated slope of $1.1$,
we divided the entropy by ${\it S} \propto r^{1.1}$
as shown in figure~\ref{fig:discussion1}(b). There appears to be a
deviation from the numerical simulation in the range of $r > 0.2\; r_{200}$,
indicating the flattening of the entropy profiles.
We note that the flattening is common to three clusters

We compare our result with a hydrodynamical simulation by
\citet{takizawa98} that allowed for different electron and ion
temperatures. We fit a $\beta$-model density profile (parameters $n_{0},
r_{c}, \beta$) and a polytrope electron temperature profile (parameter
polytrope index $\gamma_{\rm p}$) using the simulated data
in his tables~1 and 2.
The resulting entropy profile shows a slope of $\gamma_{\rm p} = 0.42$
in the outer regions for the case of flat universe with
$(\Omega_0, \Lambda_0)=(0.2, 0.8)$. Even though this result might be
an extreme case, it shows that a difference in the electron and
ion temperatures can cause a flattening of the entropy profile.

\subsection{Equilibration timescale}

Ions carry most of the kinetic energy in the cluster outskirts, and
they will be thermalized fairly quickly after accretion shocks or
mergers.  However, heating the electrons takes a long time because of
the inefficient energy transfer between ions and electrons;
the equilibration time for electron-ion collisions ($t_{\rm ei}$) is about
2000 times longer than electron-electron process ($t_{\rm ee}$) and about
45 times longer than ion-ion relaxation time ($t_{\rm ii}$).

According to \citet{fox97}, \citet{takizawa98}, and \citet{rudd09},
the electron--ion timescale including contributions from both protons
and He$^{2+}$ is estimated as
\citep{spitzer56}
\begin{eqnarray}
t_{\rm ei} &\approx&
  2.0\times 10^{8}\ {\rm yr}\;
\frac{(T_{\rm e}/10^{8}~{\rm K})^{3/2}}
     {(n_{\rm i}/10^{-3}\ {\rm cm}^{-3})\; ({\ln \Lambda}/40)},
\end{eqnarray}
where $\ln \Lambda$ is the Coulomb logarithm. We simply assume that
ions are initially heated through accretion shocks at $r_{200}$. In the
post-shock region, ions achieve thermal equilibrium with a timescale
of $t_{\rm ii}$ after this heating. The ion temperature $T_{\rm i}$
will then be significantly hotter than the electron temperature
$T_{\rm e}$\@. Eventually, thermal energy is transferred from ions to
electrons through Coulomb collisions, and $T_{\rm e}$ will equal $T_{\rm i}$
after the relaxation time $t_{\rm ei}$.

\begin{figure*}[htbp]
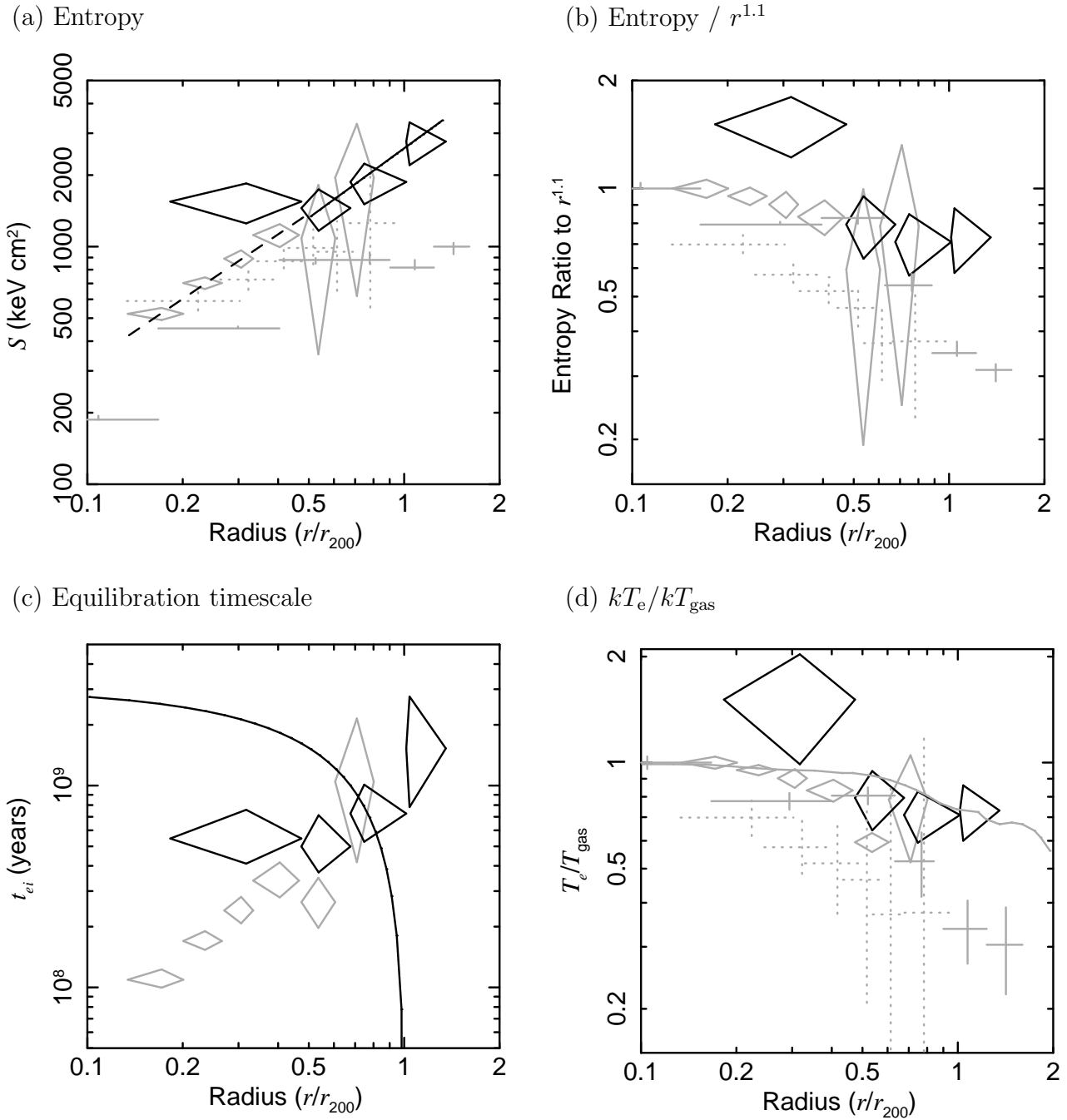

\begin{center}
\parbox{0.48\textwidth}{\ (a) Entropy}\hfill
\parbox{0.48\textwidth}{\ (b) Entropy / $r^{1.1}$}
\begin{minipage}{\textwidth}
\vspace*{-1.5ex}
\FigureFile(0.48\textwidth,\textwidth){figure8a.eps}\hfill
\FigureFile(0.48\textwidth,\textwidth){figure8b.eps}
\medskip
\end{minipage}
\parbox{0.48\textwidth}{\ (c) Equilibration timescale}\hfill
\parbox{0.48\textwidth}{\ (d) $kT_{\rm e} / kT_{\rm gas}$}
\begin{minipage}{\textwidth}
\FigureFile(0.48\textwidth,\textwidth){figure8c.eps}\hfill
\FigureFile(0.48\textwidth,\textwidth){figure8d.eps}
\end{minipage}
\end{center}
\caption{
(a) Entropy profiles (black diamond: Suzaku, grey diamond:
XMM-Newton, black solid line: fitted model to Suzaku in $7'-20'$,
black dashed line: fitted model to XMM-Newton in $0'.5-7'$,
grey solid cross: PKS0745$-$191, grey dotted cross: A1795).
(b) Entropy normalized to $\propto r^{1.1}$ profile.
(c) $t_{\rm ei}$ profile (diamonds)
compared with $t_{\rm elapsed}$ (black solid line).
(d) $T_{\rm e}/T_{\rm gas}$ profiles
compared with the simulated result by \citet{rudd09}.
}\label{fig:discussion1}
\end{figure*}

We can compare the position-dependent time since the shock heating,
$t_{\rm elapsed}$, with the equilibration timescale $t_{\rm ei}$. If
$t_{\rm ei}$ is longer than $t_{\rm elapsed}$, then $T_{\rm e}$
would be expected to be significantly lower than $T_{\rm i}$ at that position.
Denoting the velocity of inward propagation of the shock front as
$v_{\rm shock}$, we obtain
\begin{eqnarray}
r_{200} - r&\simeq& t_{\rm elapsed}\; v_{\rm shock}.
\end{eqnarray}
The free-fall velocity of the gas at $r_{200}$ is $v_{\rm ff,200} =
\sqrt{2GM_{200}/r_{200}}$.  Using the strong shock approximation and
neglecting the post-shock gas velocity compared with $v_{\rm shock}$,
\citet{takizawa98} found
\begin{eqnarray}
v_{\rm shock} &\simeq& \frac{1}{3}\; v_{\rm ff,200}.
\end{eqnarray}\label{eq:vshock}
Then, we can derive
\begin{eqnarray}
t_{\rm elapsed} &\simeq& 3\; \frac{r_{200} - r}{v_{\rm ff,200}},
\end{eqnarray}
which is independent of $M_{200}$.
In figure~\ref{fig:discussion1}(c), we show $t_{\rm elapsed}$
and $t_{\rm ei}$. 
In the region outside of $r\sim 0.9\; r_{200}$, $t_{\rm ei}$ is
significantly longer than $t_{\rm elapsed}$. Based on this
calculation, it is likely that $T_{\rm e}$ and $T_{\rm i}$ are significantly
different in the outskirts of the A1413 cluster.

\subsection{Difference between Electron and Ion Temperatures}
\citet{fox97} were the first to investigate the two-temperature nature
of the ICM\@. \citet{takizawa98} showed that in a one dimension numerical
simulation there existed a significant difference between the electron
and ion temperatures, which will affect the entropy profile and the
inferred gravitational mass. Recently, \citet{rudd09} reported the
results of simulations which indicated that the temperature difference
had a maximum of about 30\% at $r_{200}$. We will examine here a
possible deviation between electron and ion temperatures.  These
studies can help to understand how the cluster gas obtains hydrostatic
equilibrium over large volumes.

We define the average gas temperature as,
\begin{eqnarray}
T_{\rm gas} &=&
\frac{n_{\rm e} T_{\rm e} + n_{\rm i} T_{\rm i}}{n_{\rm e}+n_{\rm i}},
\end{eqnarray}
which will change over a typical electron-ion equilibration timescale,
$t_{\rm ei}$.  We estimate the average gas temperature,
$kT_{\rm gas} = S\, n_{\rm e}^{2/3}$,
by assuming a single power-law with $\gamma=1.1$
for the radial entropy profile, normalized in the cluster inner
regions where $T_{\rm i} = T_{\rm e}$ because the relaxation times are much
shorter there.  figure~\ref{fig:discussion1}(d) shows the ratio of
the observed electron temperature to the estimated average gas
temperature, where we have adopted $n_{\rm i} = 0.92\; n_{\rm e}$
(including He$^{2+}$) for a fully ionized gas
with $X = 0.7$ and $Y = 0.28$. Temperature inconsistency
between $T_{\rm e}$ and $T_{\rm gas}$ is possibly larger than
the simulation example \citep{rudd09}.

The rapid $T_{\rm e}$ decrease in the cluster outer regions may be
explained by either the ICM not being in hydrostatic equilibrium or by
differences between $T_{\rm e}$ and $T_{\rm i}$. We could determine which
interpretation is correct if we could directly estimate $T_{\rm i}$ from the
line width. This measurement should be possible in the near future
using the microcalorimeters on the ASTRO-H mission \citep{takahashi08}.

\subsection{Mass Estimation to $r_{200}$}
We calculated the gravitational mass of A1413 to $r_{200}$ assuming spherical
symmetry and hydrostatic equilibrium. From numerical simulations, these
assumptions are valid within $\sim 2\; r_{200}$ except for the core region
at $r < 0.3\; r_{200}$, where cooling and heating of AGN give significant
effects on the physical state of the gas \citep{roncarelli06,borgani06}.
Previous X-ray studies mainly showed gravitational mass within
$r_{500}$ because of instrumental limitations.
In this section, we determine the mass profile in the outer region of
A1413.

Assuming hydrostatic equilibrium, the total integrated gravitational
mass, $M_{<R}$, within the 3-dimensional radius $R$ is given by
\citep{fabricant80}
\begin{eqnarray}
M_{<R} &=& -\frac{R^{2}}{\rho_g G} \frac{d P_g}{d R} \\
  &=& -\frac{kTR}{\mu m_p G}\left(\frac{d\ln\rho_g}{d\ln R}+ 
\frac{d\ln T}{d\ln R} \right).
\label{eq:mass}
\end{eqnarray}
where $G$ is the gravitational constant, $\mu$ is the mean molecular
weight of the gas and $m_p$ is the proton mass.
We derive the above temperature and gas density profiles using the
observed projected temperature and surface brightness profiles. We use
the projected temperature directly, but discuss the validity of this
assumption below.  We calculate the gas density from the normalization
of the ICM spectral fit by taking into account the projection effect.
The {\it apec} normalization parameter is defined as
${\it Norm} = 10^{-14}
\int n_{\rm e} n_{\rm H} dV / (4\pi(1+z)^{2} D_{\rm A}^2)$~cm$^{-5}$,
with $D_{\rm A}$ the angular diameter distance to the source.
We estimated the de-projected $n_{\rm e} n_{\rm H}$ values assuming
spherical symmetry and a constant temperature in each annular
region as described in Appendix 1, and then assumed
$n_{\rm e} = 1.2\; n_{\rm H}$ (excluding He$^{2+}$) as described above.

\begin{figure*}[bt]
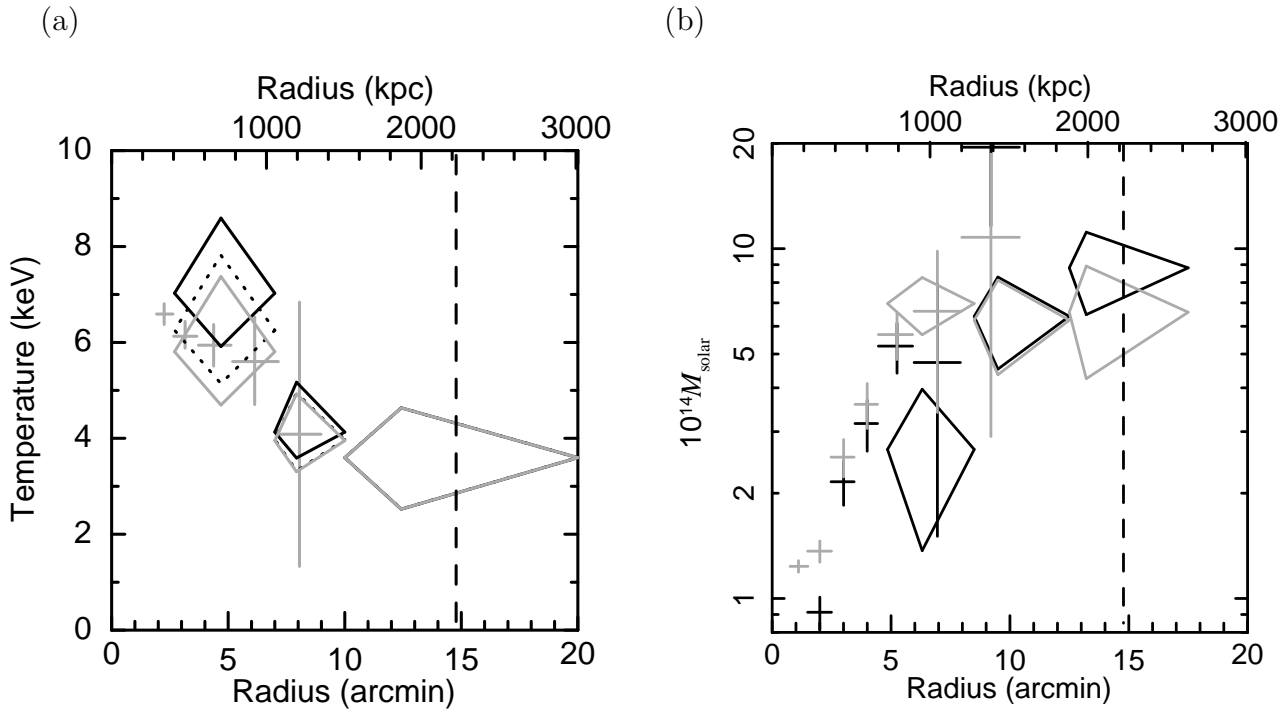

\begin{center}
\parbox{0.48\textwidth}{\ (a) }
\parbox{0.48\textwidth}{\ (b) }
\begin{minipage}{\textwidth}
\FigureFile(0.48\textwidth,\textwidth){figure9a.eps}\hfill
\FigureFile(0.48\textwidth,\textwidth){figure9b.eps}
\end{minipage}
\end{center}
\caption{
(a)Comparison of observed temperature (black diamonds) with $kT_{\rm ew}$ (doted diamonds), and $kT_{\rm sl}$ (grey diamonds). Grey crosses show $kT_{\rm 3d}$ observed with XMM-Newton by \citet{snowden08}. (b)Integrated mass profile (black diamonds: Suzaku with $T_{\rm gas}$, grey diamonds: Suzaku with $T_{\rm e}$,
black crosses: XMM-Newton with $T_{\rm gas}$,
and grey crosses: XMM-Newton with $T_{\rm e}$).
Vertical dashed
line shows $r_{200}=14'.8$.
}\label{fig:discussion2}
\end{figure*}

Allowing for the possibility of $T_{\rm e} \neq T_{\rm i}$, we consider two
cases for $T$: the electron temperature and the average gas
temperature.  We show the integrated mass profiles in
figure~\ref{fig:discussion2}(a) based on $kT_{\rm e}$ and $kT_{\rm gas}$.
These profiles are obtained without using any particular model since we
perform the needed derivatives by differencing the temperatures and
densities of adjacent radial bins. The integrated mass within
$13'.2_{-0'.7}^{+4'.3}$, which encompasses $r_{200}$ ($14'.8$) is
$(8.8\pm 2.3) \times 10^{14} M_{\odot}$ using $kT_{\rm gas}$. This mass is
about 30\% larger than that obtained using $kT_{\rm e}$ of
$(6.6\pm 2.3)\times 10^{14} M_{\odot}$, although the difference is not
statistically significant.  The 30\% difference in the temperatures
propagates almost directly to the same mass difference. Our mass
determination agrees with that of \citet{vikhlinin06}, but not with
\citet{pointecouteau05}. These masses imply an overdensity with
respect to critical of 177$\pm$47 and 132$\pm$47, where the errors
are only from the mass errors.

In the above mass estimation, we assumed that the observed projected
temperature is the 3-dimensional value at the observed radius.  We
need to examine the systematic error caused by this assumption.  In
the following we denote the true 3-dimensional temperature of the ICM
by $T_{\rm 3d}$, which varies with radius.  We derive the temperature 
from the spectral fit is a weighted mean of different temperatures
projected along the line of sight. Often the projected temperature is
defined as the emission-weighted temperature $T_{\rm ew}$,
\begin{eqnarray}
T_{\rm ew}\equiv\frac{\int n^2 \Lambda (T) T dV}{\int n^2 \Lambda(T) dV}\ .
\end{eqnarray}
However, \citet{mazzotta04} discussed how the spectral response of an
actual instrument implies that $T_{\rm ew}$ can be quite different from
what would be measured with that instrument observing a non-isothermal
temperature distribution.  For a better approximation, they introduced
a spectroscopic-like temperature $T_{\rm sl}$ defined as,
\begin{eqnarray}
   T_{\rm sl}\equiv\frac{\int n^2 T^{a-1/2} dV}{\int n^2 T^{a-3/2} dV}\quad,
\end{eqnarray}
with $a=0.75$, which empirically gave a good estimate of the $T$
measured with XMM-Newton or Chandra.  \citet{rasia05} and
\citet{shimizu06} reported that the difference between $T_{\rm ew}$
and $T_{\rm sl}$ can be as large as 30\%. We carried out comparison of
observed temperature with $kT_{\rm ew}$ and $kT_{\rm sl}$ in figure
\ref{fig:discussion2}(a). The difference between $kT_{\rm ew}$ and
$kT_{\rm sl}$ takes the largest value of about 8.2\% in the radius
$2'.6-7'.0$. These temperatures are consistent with the observed data
with XMM-Newton.  Taking a conservative value, our mass estimate would
be more than 30\% different from the true value because of our
employment of the observed projected temperature as the 3-dimensional
one.

\section{Summary}

\begin{itemize}
\setlength{\itemsep}{1ex}

\item
Northern outskirts of the relaxed cluster of galaxies A1413 was
observed with Suzaku in the radial range of $2.'7-26'$ covering
the virial radius of $r_{200}= 14'.8$.
We excised 15 point sources above a flux of
$\sim 1\times 10^{-14}$ \fluxunit\ (2--10 keV),
and the CXB level after the point source excision was evaluated.
We quantify all known systematic errors,
and show statistical errors are dominant.

\item
Suzaku detected X-ray emission of the ICM up to the $15'-20'$ annulus
beyond the virial radius.
Significant temperature decrease to $\sim 3$~keV (factor of $\sim 2$)
at $r_{200}$ is confirmed, which was reported in a few other
clusters, PKS0745$-$191 \citep{george08}, A1795 \citep{bautz09},
and A2204 \citep{reiprich09}.

\ifnum1=0
\item
We tried to explain our measured temperature and surface brightness
with SSM model which is cluster model with 6 parameters assuming
spherical symmetry and hydrostatistic equilibrium. We found that,
although we could fit either the temperature or surface brightness
profile, it was not possible to simultaneously fit both despite an
exhaustive search of parameter space. The likely reason for this
result is that the ICM is out of equilibrium in the outer regions of
the cluster.
\fi

\item
Our entropy profile in the outer region ($> 0.5\; r_{200}$) joins
smoothly onto that of XMM-Newton at 0.15--$0.5\; r_{200}$,
and shows a flatter slope of $\gamma=0.90\pm 0.12$ than
$\gamma=1.1$ \citep{voit05} obtained with numerical simulations of
adiabatic gas accretion.

\item Deviation of the entropy profile from the $r^{-1.1}$ relation
  would show that electron temperature is not equal to gas temperature
  in outer region, where equilibration timescale for electron-ion
  collision, $t_{\rm ei}$, is longer than the elapsed time after the
  shock heating, $t_{\rm elapsed}$.

\item
The integrated mass of the cluster at the virial radius is
approximately $7.5\times 10^{14}\; M_{\odot}$ and varies by $\sim 30$\%
depending on temperatures ($T_{\rm e}$, $T_{\rm gas}$,
$T_{\rm ew}$, and $T_{\rm sl}$) which we use.

\end{itemize}

\bigskip
We are grateful to S.~L.~Snowden and A.~Vikhlinin for communicating
their unpublished results.  This work was supported by Grant-in-Aid
for JSPS Fellows (20$\cdot$2427) and the MEXT program ``Support
Program for Improving Graduate School Education''. JPH gratefully
acknowledges financial support from NASA grant NNG06GC04G.

\ifnum1=0
\section{Relation Between the Emission Measure and the APEC Model 
Normalization}
The relation between the emission measure and the emission integral is
\begin{eqnarray}\label{eq:emei}
\frac{\it SOURCE\_RATIO\_REG}{\Omega_{\rm e}}\int n_{\rm e} n_{\rm H} dV =
\int n_{\rm e} n_{\rm H} dl
\hspace*{1em}{\rm (cm^{-3}~kpc^{-2})}
\end{eqnarray}
The normalization of the APEC spectral model is
\begin{eqnarray}\label{eq:norm}
{\it Norm} = \frac{10^{-14}}{4\pi D_{\rm A}^2\left(1+z\right)^2}\int n_{\rm e} n_{\rm H} dV   
\hspace*{1em}{\rm (cm^{-5})}
\end{eqnarray}
From equation (\ref{eq:emei}) and equation (\ref{eq:norm}),
\begin{eqnarray}
{\it Norm} = \frac{10^{-14}}{4\pi D_{\rm A}^2\left(1+z\right)^2}\frac{\Omega_{\rm e}}{\it SOURCE\_RATIO\_REG}\int n_{\rm e} n_{\rm H} dl
\hspace*{1em}{\rm (cm^{-5})}.
\end{eqnarray}
\fi

\end{document}